\documentclass[letterpaper,compsoc,twoside]{IEEEtran}
\usepackage{fixltx2e} 
\usepackage{cmap} 
\usepackage{ifthen}
\usepackage[T1]{fontenc}
\usepackage[utf8]{inputenc}
\usepackage{amsmath}

\usepackage[font={small,it},labelfont=bf]{caption}
\usepackage{float}

\pdfoutput=1
\usepackage{scipy}
\makeatletter
\def\PY@reset{\let\PY@it=\relax \let\PY@bf=\relax%
    \let\PY@ul=\relax \let\PY@tc=\relax%
    \let\PY@bc=\relax \let\PY@ff=\relax}
\def\PY@tok#1{\csname PY@tok@#1\endcsname}
\def\PY@toks#1+{\ifx\relax#1\empty\else%
    \PY@tok{#1}\expandafter\PY@toks\fi}
\def\PY@do#1{\PY@bc{\PY@tc{\PY@ul{%
    \PY@it{\PY@bf{\PY@ff{#1}}}}}}}
\def\PY#1#2{\PY@reset\PY@toks#1+\relax+\PY@do{#2}}

\expandafter\def\csname PY@tok@gd\endcsname{\def\PY@tc##1{\textcolor[rgb]{0.63,0.00,0.00}{##1}}}
\expandafter\def\csname PY@tok@gu\endcsname{\let\PY@bf=\textbf\def\PY@tc##1{\textcolor[rgb]{0.50,0.00,0.50}{##1}}}
\expandafter\def\csname PY@tok@gt\endcsname{\def\PY@tc##1{\textcolor[rgb]{0.00,0.27,0.87}{##1}}}
\expandafter\def\csname PY@tok@gs\endcsname{\let\PY@bf=\textbf}
\expandafter\def\csname PY@tok@gr\endcsname{\def\PY@tc##1{\textcolor[rgb]{1.00,0.00,0.00}{##1}}}
\expandafter\def\csname PY@tok@cm\endcsname{\let\PY@it=\textit\def\PY@tc##1{\textcolor[rgb]{0.25,0.50,0.56}{##1}}}
\expandafter\def\csname PY@tok@vg\endcsname{\def\PY@tc##1{\textcolor[rgb]{0.73,0.38,0.84}{##1}}}
\expandafter\def\csname PY@tok@m\endcsname{\def\PY@tc##1{\textcolor[rgb]{0.13,0.50,0.31}{##1}}}
\expandafter\def\csname PY@tok@mh\endcsname{\def\PY@tc##1{\textcolor[rgb]{0.13,0.50,0.31}{##1}}}
\expandafter\def\csname PY@tok@cs\endcsname{\def\PY@tc##1{\textcolor[rgb]{0.25,0.50,0.56}{##1}}\def\PY@bc##1{\setlength{\fboxsep}{0pt}\colorbox[rgb]{1.00,0.94,0.94}{\strut ##1}}}
\expandafter\def\csname PY@tok@ge\endcsname{\let\PY@it=\textit}
\expandafter\def\csname PY@tok@vc\endcsname{\def\PY@tc##1{\textcolor[rgb]{0.73,0.38,0.84}{##1}}}
\expandafter\def\csname PY@tok@il\endcsname{\def\PY@tc##1{\textcolor[rgb]{0.13,0.50,0.31}{##1}}}
\expandafter\def\csname PY@tok@go\endcsname{\def\PY@tc##1{\textcolor[rgb]{0.20,0.20,0.20}{##1}}}
\expandafter\def\csname PY@tok@cp\endcsname{\def\PY@tc##1{\textcolor[rgb]{0.00,0.44,0.13}{##1}}}
\expandafter\def\csname PY@tok@gi\endcsname{\def\PY@tc##1{\textcolor[rgb]{0.00,0.63,0.00}{##1}}}
\expandafter\def\csname PY@tok@gh\endcsname{\let\PY@bf=\textbf\def\PY@tc##1{\textcolor[rgb]{0.00,0.00,0.50}{##1}}}
\expandafter\def\csname PY@tok@ni\endcsname{\let\PY@bf=\textbf\def\PY@tc##1{\textcolor[rgb]{0.84,0.33,0.22}{##1}}}
\expandafter\def\csname PY@tok@nl\endcsname{\let\PY@bf=\textbf\def\PY@tc##1{\textcolor[rgb]{0.00,0.13,0.44}{##1}}}
\expandafter\def\csname PY@tok@nn\endcsname{\let\PY@bf=\textbf\def\PY@tc##1{\textcolor[rgb]{0.05,0.52,0.71}{##1}}}
\expandafter\def\csname PY@tok@no\endcsname{\def\PY@tc##1{\textcolor[rgb]{0.38,0.68,0.84}{##1}}}
\expandafter\def\csname PY@tok@na\endcsname{\def\PY@tc##1{\textcolor[rgb]{0.25,0.44,0.63}{##1}}}
\expandafter\def\csname PY@tok@nb\endcsname{\def\PY@tc##1{\textcolor[rgb]{0.00,0.44,0.13}{##1}}}
\expandafter\def\csname PY@tok@nc\endcsname{\let\PY@bf=\textbf\def\PY@tc##1{\textcolor[rgb]{0.05,0.52,0.71}{##1}}}
\expandafter\def\csname PY@tok@nd\endcsname{\let\PY@bf=\textbf\def\PY@tc##1{\textcolor[rgb]{0.33,0.33,0.33}{##1}}}
\expandafter\def\csname PY@tok@ne\endcsname{\def\PY@tc##1{\textcolor[rgb]{0.00,0.44,0.13}{##1}}}
\expandafter\def\csname PY@tok@nf\endcsname{\def\PY@tc##1{\textcolor[rgb]{0.02,0.16,0.49}{##1}}}
\expandafter\def\csname PY@tok@si\endcsname{\let\PY@it=\textit\def\PY@tc##1{\textcolor[rgb]{0.44,0.63,0.82}{##1}}}
\expandafter\def\csname PY@tok@s2\endcsname{\def\PY@tc##1{\textcolor[rgb]{0.25,0.44,0.63}{##1}}}
\expandafter\def\csname PY@tok@vi\endcsname{\def\PY@tc##1{\textcolor[rgb]{0.73,0.38,0.84}{##1}}}
\expandafter\def\csname PY@tok@nt\endcsname{\let\PY@bf=\textbf\def\PY@tc##1{\textcolor[rgb]{0.02,0.16,0.45}{##1}}}
\expandafter\def\csname PY@tok@nv\endcsname{\def\PY@tc##1{\textcolor[rgb]{0.73,0.38,0.84}{##1}}}
\expandafter\def\csname PY@tok@s1\endcsname{\def\PY@tc##1{\textcolor[rgb]{0.25,0.44,0.63}{##1}}}
\expandafter\def\csname PY@tok@gp\endcsname{\let\PY@bf=\textbf\def\PY@tc##1{\textcolor[rgb]{0.78,0.36,0.04}{##1}}}
\expandafter\def\csname PY@tok@sh\endcsname{\def\PY@tc##1{\textcolor[rgb]{0.25,0.44,0.63}{##1}}}
\expandafter\def\csname PY@tok@ow\endcsname{\let\PY@bf=\textbf\def\PY@tc##1{\textcolor[rgb]{0.00,0.44,0.13}{##1}}}
\expandafter\def\csname PY@tok@sx\endcsname{\def\PY@tc##1{\textcolor[rgb]{0.78,0.36,0.04}{##1}}}
\expandafter\def\csname PY@tok@bp\endcsname{\def\PY@tc##1{\textcolor[rgb]{0.00,0.44,0.13}{##1}}}
\expandafter\def\csname PY@tok@c1\endcsname{\let\PY@it=\textit\def\PY@tc##1{\textcolor[rgb]{0.25,0.50,0.56}{##1}}}
\expandafter\def\csname PY@tok@kc\endcsname{\let\PY@bf=\textbf\def\PY@tc##1{\textcolor[rgb]{0.00,0.44,0.13}{##1}}}
\expandafter\def\csname PY@tok@c\endcsname{\let\PY@it=\textit\def\PY@tc##1{\textcolor[rgb]{0.25,0.50,0.56}{##1}}}
\expandafter\def\csname PY@tok@mf\endcsname{\def\PY@tc##1{\textcolor[rgb]{0.13,0.50,0.31}{##1}}}
\expandafter\def\csname PY@tok@err\endcsname{\def\PY@bc##1{\setlength{\fboxsep}{0pt}\fcolorbox[rgb]{1.00,0.00,0.00}{1,1,1}{\strut ##1}}}
\expandafter\def\csname PY@tok@kd\endcsname{\let\PY@bf=\textbf\def\PY@tc##1{\textcolor[rgb]{0.00,0.44,0.13}{##1}}}
\expandafter\def\csname PY@tok@ss\endcsname{\def\PY@tc##1{\textcolor[rgb]{0.32,0.47,0.09}{##1}}}
\expandafter\def\csname PY@tok@sr\endcsname{\def\PY@tc##1{\textcolor[rgb]{0.14,0.33,0.53}{##1}}}
\expandafter\def\csname PY@tok@mo\endcsname{\def\PY@tc##1{\textcolor[rgb]{0.13,0.50,0.31}{##1}}}
\expandafter\def\csname PY@tok@mi\endcsname{\def\PY@tc##1{\textcolor[rgb]{0.13,0.50,0.31}{##1}}}
\expandafter\def\csname PY@tok@kn\endcsname{\let\PY@bf=\textbf\def\PY@tc##1{\textcolor[rgb]{0.00,0.44,0.13}{##1}}}
\expandafter\def\csname PY@tok@o\endcsname{\def\PY@tc##1{\textcolor[rgb]{0.40,0.40,0.40}{##1}}}
\expandafter\def\csname PY@tok@kr\endcsname{\let\PY@bf=\textbf\def\PY@tc##1{\textcolor[rgb]{0.00,0.44,0.13}{##1}}}
\expandafter\def\csname PY@tok@s\endcsname{\def\PY@tc##1{\textcolor[rgb]{0.25,0.44,0.63}{##1}}}
\expandafter\def\csname PY@tok@kp\endcsname{\def\PY@tc##1{\textcolor[rgb]{0.00,0.44,0.13}{##1}}}
\expandafter\def\csname PY@tok@w\endcsname{\def\PY@tc##1{\textcolor[rgb]{0.73,0.73,0.73}{##1}}}
\expandafter\def\csname PY@tok@kt\endcsname{\def\PY@tc##1{\textcolor[rgb]{0.56,0.13,0.00}{##1}}}
\expandafter\def\csname PY@tok@sc\endcsname{\def\PY@tc##1{\textcolor[rgb]{0.25,0.44,0.63}{##1}}}
\expandafter\def\csname PY@tok@sb\endcsname{\def\PY@tc##1{\textcolor[rgb]{0.25,0.44,0.63}{##1}}}
\expandafter\def\csname PY@tok@k\endcsname{\let\PY@bf=\textbf\def\PY@tc##1{\textcolor[rgb]{0.00,0.44,0.13}{##1}}}
\expandafter\def\csname PY@tok@se\endcsname{\let\PY@bf=\textbf\def\PY@tc##1{\textcolor[rgb]{0.25,0.44,0.63}{##1}}}
\expandafter\def\csname PY@tok@sd\endcsname{\let\PY@it=\textit\def\PY@tc##1{\textcolor[rgb]{0.25,0.44,0.63}{##1}}}

\makeatother

\providecommand*{\DUrole}[2]{%
  \ifcsname DUrole#1\endcsname%
    \csname DUrole#1\endcsname{#2}%
  \else
    \ifcsname docutilsrole#1\endcsname%
      \csname docutilsrole#1\endcsname{#2}%
    \else%
      #2%
    \fi%
  \fi%
}

\ifthenelse{\isundefined{\hypersetup}}{
  \usepackage[colorlinks=true,linkcolor=blue,urlcolor=blue]{hyperref}
  \urlstyle{same} 
}{}

\begin{document}
\newcounter{footnotecounter}\title{Temperature diagnostics of the solar atmosphere using SunPy}\author{Andrew Leonard$^{\setcounter{footnotecounter}{1}\fnsymbol{footnotecounter}\setcounter{footnotecounter}{2}\fnsymbol{footnotecounter}}$%
          \setcounter{footnotecounter}{1}\thanks{\fnsymbol{footnotecounter} %
          Corresponding author: \protect\href{mailto:ajl7@aber.ac.uk}{ajl7@aber.ac.uk}}\setcounter{footnotecounter}{2}\thanks{\fnsymbol{footnotecounter} Institute of Mathematics, Physics and Computer Science, Aberystwyth University, Ceredigion, SY23 3BZ, Wales}, Huw Morgan$^{\setcounter{footnotecounter}{2}\fnsymbol{footnotecounter}}$\thanks{%

          \noindent%
          Copyright\,\copyright\,2014 Andrew Leonard et al. This is an open-access article distributed under the terms of the Creative Commons Attribution License, which permits unrestricted use, distribution, and reproduction in any medium, provided the original author and source are credited. http://creativecommons.org/licenses/by/3.0/%
        }}\maketitle
          \renewcommand{\leftmark}{PROC. OF THE 7th EUR. CONF. ON PYTHON IN SCIENCE (EUROSCIPY 2014)}
          \renewcommand{\rightmark}{TEMPERATURE DIAGNOSTICS OF THE SOLAR ATMOSPHERE USING SUNPY}

\setcounter{page}{11}
\newcommand*{\docutilsroleref}{\ref}
\newcommand*{\docutilsrolelabel}{\label}
\AtEndDocument{\cleardoublepage}
\begin{abstract}The solar atmosphere is a hot (\textasciitilde{}1MK), magnetised plasma of great
interest to physicists. There have been many previous studies of the
temperature of the Sun's atmosphere (\cite{Plowman2012}, \cite{Wit2012},
\cite{Hannah2012}, \cite{Aschwanden2013}, etc.). Almost all of these studies use
the SolarSoft software package written in the commercial Interactive Data
Language (IDL), which has been the standard language for solar physics.
The SunPy project aims to provide an open-source library for solar physics.
This work presents (to the authors' knowledge) the first study of its type
to use SunPy rather than SolarSoft.

This work uses SunPy to process multi-wavelength solar observations made by
the Atmospheric Imaging Assembly (AIA) instrument aboard the Solar Dynamics
Observatory (SDO) and produce temperature maps of the Sun's atmosphere. The
method uses SunPy's utilities for querying databases of solar events,
downloading solar image data, storing and processing images as spatially
aware Map objects, and tracking solar features as the Sun rotates. An
essential consideration in developing this software is computational
efficiency due to the large amount of data collected by AIA/SDO, and in
anticipating new solar missions which will result in even larger sets of
data. An overview of the method and implementation is given, along with
tests involving synthetic data and examples of results using real data for
various regions in the Sun's atmosphere.\end{abstract}\begin{IEEEkeywords}solar, corona, data mining, image processing\end{IEEEkeywords}

\section{Introduction%
  \label{introduction}%
}

The solar corona is a hot (\textasciitilde{}1MK) magnetised plasma. Such an environment,
difficult to reproduce in a laboratory, is of great importance in physics (e.g.
basic plasma physics, development of nuclear fusion). It is important also in
the context of general astronomy, in understanding other stars and the
mechanisms which heat the corona - considered one of the major unanswered
questions in astronomy. On a more practical level, with our society's
growing reliance on space-based technology, we are increasingly prone to the
effects of geo-effective solar phenomena such as flares and Coronal Mass
Ejections (CMEs). These can damage the electronic infrastructure which plays an
essential role in our modern society. In order to be able to predict these
phenomena, we must first understand their formation and development. The
sources of energy for these events can cause localised heating in the corona,
and coronal temperature distributions are therefore a widely studied topic
within solar physics. Active regions, which are regions of newly-emerged
magnetic flux from the solar interior associated with sunspots, are of
particular interest since they are often the source regions for the most
damaging eruptive events and the distribution of temperatures can give us
unique and valuable information on the initial conditions of eruptions.

It is not yet technologically feasible to send probes into the low coronal
environment. Our current understanding of the corona is based on remote-sensing
observations across the electromagnetic spectrum from radio to X-ray. Previous
studies have found coronal temperatures ranging from \textasciitilde{}0.8MK in coronal hole
regions, \textasciitilde{}1MK in quiet Sun regions and from 1-3MK within active regions.
Eruptive events and flares can produce even higher temperatures in small
regions for a short period of time (see, e.g., \cite{Awasthi2014}). This work is
based exclusively on images of the low corona taken in Extreme Ultra-Violet
(EUV) using AIA/SDO \cite{Lemen2011}. AIA allows us, for the first time, to produce
reliable maps of the coronal temperature with very fine spatial and temporal
resolution.

An overview of the AIA instrument is given in Section \DUrole{ref}{instmeth}, along
with an introduction to the theory of estimating temperature from spectral
observation. The method is tested by creating synthetic AIA data created from a
coronal emission model in Section \DUrole{ref}{modeltests}. The method is applied to
AIA data in Section \DUrole{ref}{results} and the results compared to those of
previous studies. Discussion and Conclusions are given in Section \DUrole{ref}{disc}.

\section{Instrumentation \& Method%
  \label{instrumentation-method}%
}

\DUrole{label}{instmeth}

\subsection{Overview of AIA/SDO%
  \label{overview-of-aia-sdo}%
}

The AIA on NASA's SDO satellite \cite{Lemen2011} was launched in February 2010 and
started making regular observations in March 2010. AIA takes a 4096 x 4096
full-disk image of the corona in each of ten wavelength channels every \textasciitilde{}12
seconds, with a resolution of \textasciitilde{}1.2 arcsec per pixel. Seven of these channels
observe within the EUV wavelength range, of which six are each dominated by
emission from a different Fe ion. Figure \DUrole{ref}{exampleAIAimage} shows an
example AIA image using the 17.1nm channel. It has been processed to enhance
smaller-scale features using a newly-developed method \cite{Morgan2014}. The
intensity measured in each of the six narrow-band Fe channels (9.4nm, 13.1nm,
17.1nm, 19.3nm, 21.1nm and 33.5nm) is dominated by a spectral line of iron at a
specific stage of ionisation. Simply speaking, therefore, they correspond to
different temperatures of the emitting plasma. The fact that there are six
channels observing simultaneously means that the temperature of the corona can
be effectively constrained \cite{Guennou2012}. Another advantage is that relative
elemental abundances do not need to be considered when using emission from only
one element, thus reducing the associated errors. AIA is therefore beginning to
be widely used for this type of study (e.g. \cite{Aschwanden2013}), as its very
high spatial, temporal and thermal resolution make it an excellent source of
data for investigating the temperatures of small-scale and/or dynamic features
in the corona, as well as for looking at global and long-term temperature
distributions.

It is important that the large amount of data produced by AIA
can be analysed quickly. A way of calculating coronal temperatures in real-time
or near real-time would be extremely useful as it would allow temperature maps
to be produced from AIA images as they are taken. Computational efficiency was
therefore one of our most important criteria when designing the method for use
with AIA data.\begin{figure}[]\noindent\makebox[\columnwidth][c]{\includegraphics[width=\columnwidth]{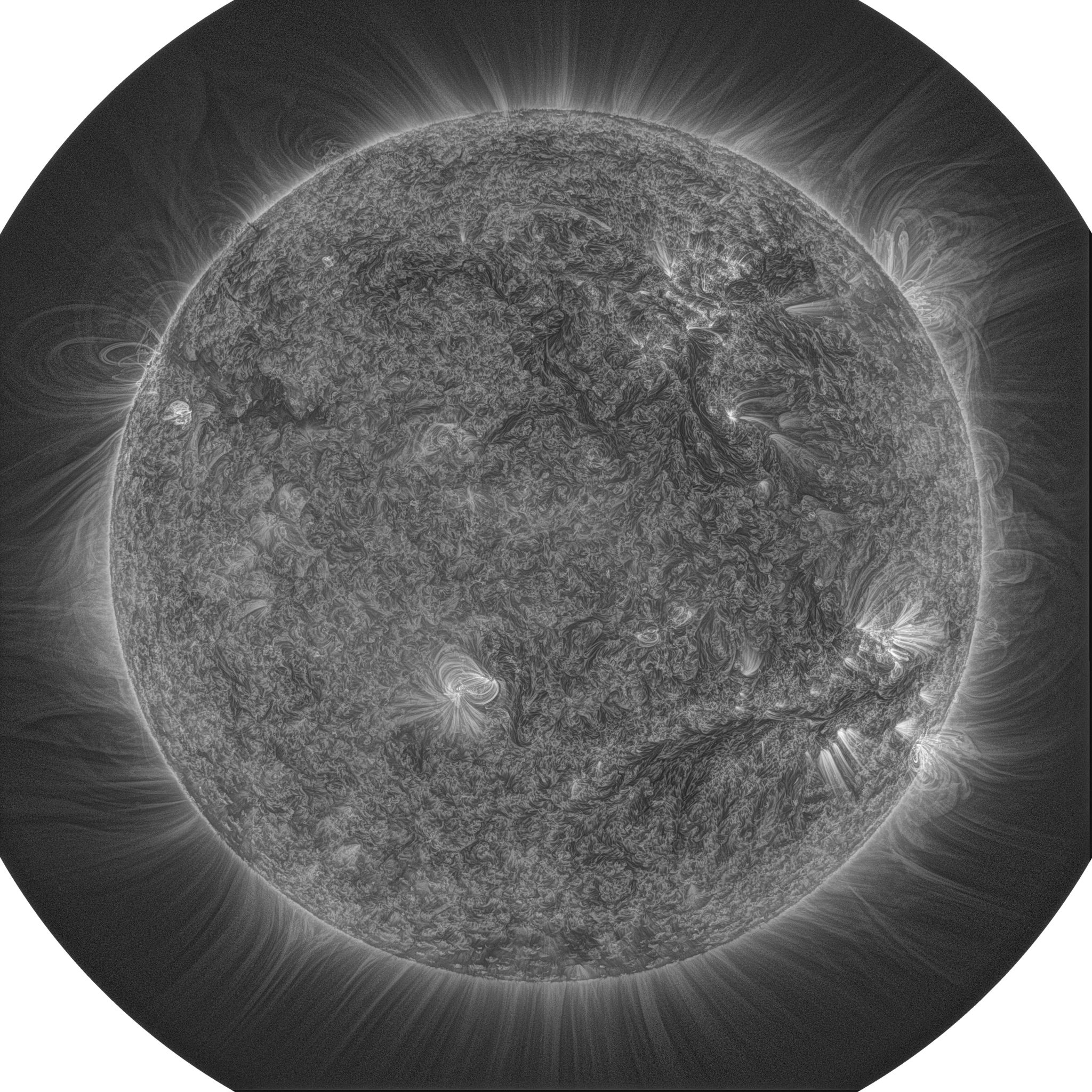}}
\caption{A typical image from the 17.1nm channel of AIA, which has been enhanced to
show small-scale features more clearly.
\DUrole{label}{exampleAIAimage}}
\end{figure}

\subsection{The Differential Emission Measure%
  \label{the-differential-emission-measure}%
}

Coronal emission lines originate from a wide range of ions which form at
different temperatures. By using multi-wavelength observations of the corona to
compare the brightnesses of the emission due to these ions, one can infer the
temperature of the corona at the location of the emission. Since the plasma may
have a range of temperatures rather than being isothermal, it is common to
describe the amount of plasma emitting along a given line-of-sight (LOS) as a
function of temperature. This function is called the Differential Emission
Measure (DEM). The DEM is usally expressed in terms of the electron density,
$n_e$ (which is not known unless already determined by some other method):\begin{equation*}
\textrm{DEM}(T)=n_{e}^{2}\frac{\textrm{d}z}{\textrm{d}T}
\end{equation*}where $z$ is the distance along the LOS and $T$ is electron
temperature.
Determining the DEM therefore gives us an estimate of the column electron
density. The width of the DEM provides a measure of how multi-thermal the
plasma is. The temperature of peak of the DEM is the dominant temperature, i.e.:
the temperature of the majority of the plasma.

The intensity measured by pixel $x$ of a particular channel $i$ on
an instrument can be expressed as a convolution of the DEM and the temperature
response function $K_{i}$ of the instrument:\begin{equation}
\label{pixelval}
I_{i}(x)=\int_{0}^{\infty}K_{i}(T)\,\textrm{DEM}(T,x)\,\textrm{d}T
\end{equation}The temperature response combines the wavelength response of the instrument and
the contribution function, which describes the emission of the plasma at a
given temperature based on atomic physics models. Unfortuately,
(\DUrole{ref}{pixelval}) is an ill-posed problem and as such there exists no unique
solution without imposing physical contraints \cite{Judge1997}. Multiple schemes
have been designed to invert this equation and infer the DEM by applying
various physical assumptions. However, these assumptions are sometimes
difficult to justify and the accuracy of the results is also reduced by the
typically high errors on solar measurements. The physical constraints assumed
by this method are discussed in Section \DUrole{ref}{DEMfinding}.

This work presents an extremely fast method of estimating the temperature of
coronal plasma from AIA images. This method is implemented using the
SunPy solar physics library (\href{http://www.sunpy.org/}{www.sunpy.org}) and produces results comparable to
those of other methods but in a fraction of the time. The current
implementation of the method is designed primarily with efficiency in mind.

\subsection{Preprocessing%
  \label{preprocessing}%
}

Level 1.0 AIA data were obtained using SunPy's wrappers around the Virtual
Solar Observatory. These data were corrected for exposure time and further
processed to level 1.5. This extra level of processing provides the correct
spatial co-alignment necessary for a quantitative comparison of the different
channels. To this end, the AIA images used were processed using the SunPy
\texttt{aiaprep()} function to ensure that all images used were properly rescaled
and co-aligned. \texttt{aiaprep()} rotates the images so that solar north points to
the top of the image, scales them so that each pixel is exactly 0.6 arcsec
across (in both the x and y directions), and recentres them so that solar
centre coincides with the centre of the image. This is achieved using an affine
transform and bi-cubic interpolation. All images were then normalised by
dividing the intensity measured in each pixel by the intensity in the
corresponding pixel in the 17.1nm image. The 17.1nm image was therefore 1 in
all pixels, and the images from all other channels are given as a ratio of the
17.1nm intensity.

\subsection{Temperature response functions%
  \label{temperature-response-functions}%
}

Temperature response functions can be calculated for each of the AIA channels
using the equation:\begin{equation}
\label{temp_response}
K_{i}(\mathrm{T})=\int_{0}^{\infty}G(\lambda,\mathrm{T})\, R_{i}(\lambda)\,\mathrm{d}\lambda
\end{equation}where $\lambda$ is the wavelength, $R_{i}(\lambda)$ is the
wavelength response of each channel $i$ and $G(\lambda,\mathrm{T})$
is the contribution function describing how radiation is emitted by the coronal
plasma. For this work the AIA temperature response functions were obtained
using the IDL aia\_get\_response function (for which no equivalent exists yet in
SunPy) and an empirical correction factor of 6.7 was applied to the 9.4nm
response function for $log(T)\le 6.3$, following the work of
\cite{Aschwanden2011}. These response functions were saved and reloaded into
Python for use with this method. As with the AIA images, each of these response
functions was normalised to the 17.1nm response by dividing the value at each
temperature by the corresponding value for 17.1nm. The response functions used
in this method (before normalisation) are shown in Figure \DUrole{ref}{response-plot}.\begin{figure}[]\noindent\makebox[\columnwidth][c]{\includegraphics[width=\columnwidth]{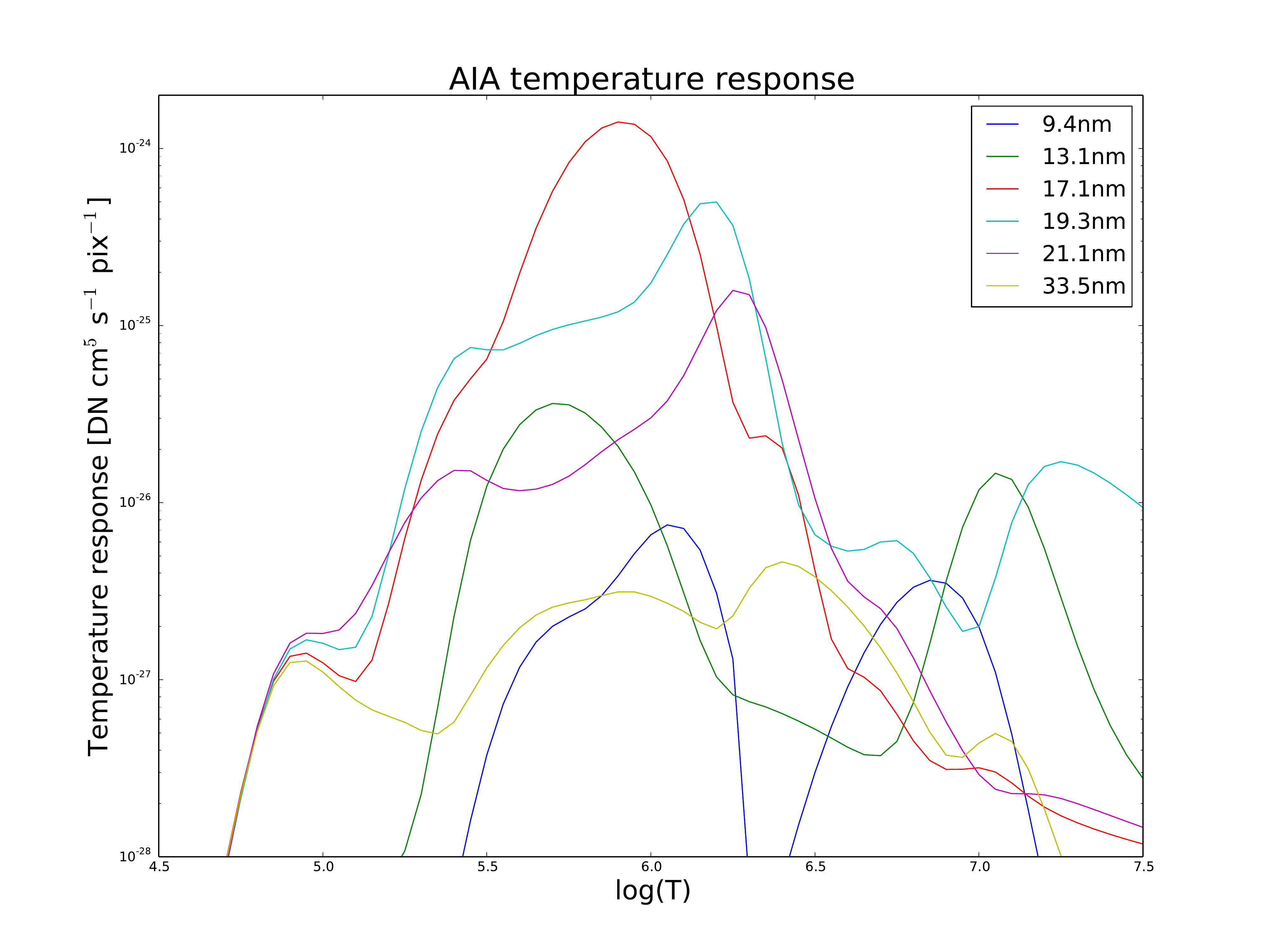}}
\caption{Temperature response of each of the Fe channels on AIA. Here the empirical
correction to the 9.4nm response has been applied but the normalisation
has not (see text). \DUrole{label}{response-plot}}
\end{figure}

\subsection{DEM-finding procedure%
  \label{dem-finding-procedure}%
}

\DUrole{label}{DEMfinding}

The general method for estimating the DEM is an iterative procedure which
systematically tests a range of possible DEMs. Each DEM is substituted into
(\DUrole{ref}{pixelval}) to produce a synthetic pixel value for each AIA
wavelength channel ($i$). This expected outcome is then compared to the
actual values measured for all pixel positions ($x$) in each wavelength,
giving a goodness-of-fit value for each pixel for a given DEM (equation
\DUrole{ref}{goodness-of-fit}), defined by the difference in pixel values averaged
over all wavelength channels:\begin{equation}
\label{goodness-of-fit}
\mathrm{fit}(x)=\frac{1}{n_{i}}\sum_{i}{|I_{measured}(x,i)-I_{synth}(x,i)|}
\end{equation}Since the synthetic emission values do not change unless one wishes to apply
different assumptions which affect the temperature response (electron density,
ionisation equilibrium, etc.), the calculation time for the method can be
reduced by saving these emission values and reusing them for each comparison.
By repeating this calculation with a number of assumed DEMs, the DEM
corresponding to the smallest goodness-of-fit value provides an estimate of the
actual plasma temperature distribution.

For this kind of iterative method to find a solution within a feasible amount
of time, a general DEM profile must be assumed. A Gaussian profile is a good
choice for the following reasons:%
\begin{itemize}

\item 

it can be fully described by only three parameters, i.e.: the mean, variance
and amplitude of the Gaussian (henceforth the peak temperature, width and
height of the DEM), which correspond to the dominant temperature, the degree
of multithermality and the peak emission measure respectively. Because
of this parameterisation, a Gaussian is well-suited to this type of method
and is also a useful way to describe important properties of the plasma even
if it does not perfectly represent the actual distribution of temperatures;
\item 

other authors have typically found multithermal DEMs, but with relatively
narrow widths (\cite{Warren2008}). \cite{Aschwanden2011} found that a narrow
Gaussian DEM fit the observations with $\chi^{2}\leq 2$ for 66\% of
cases studied, so this distribution should provide a good approximation for
the plasma in the majority of pixels. In particular, it is likely that active
region loops have a distribution of temperature and density which makes
a narrow Gaussian a physically sensible choice for the shape of the plasma
DEM. It is likely that emission from loops will dominate the measured
emission in the corresponding pixels;
\item 

since other studies have used a Gaussian DEM, using the same shape in this
work allows a direct comparison between the relative merits of the
methods themselves, without any disparity in the results caused by different
DEM profiles.
\end{itemize}

Though this particular study uses a Gaussian DEM, the method could also be used
with DEMs of any other form, such as a delta function, top hat function,
polynomial, etc. A comparison of the effect of using some of these shapes can
be found in \cite{Guennou2012a}. An active area of research is the emission of
plasma with a Kappa energy distribution—which approximates the bulk
Gaussian DEM with a high-energy population \cite{Mackovjak2014}.

The code takes a simplified approach by finding only the peak temperature of
the DEM, and assuming the height and width to be fixed. The width was set to be
0.1 and since the data are normalised relative to a given wavelength, the DEM
height is also normalised to unity. A narrow width is selected for the DEM
because, as shown by \cite{Guennou2012a}, the greater the width of the plasma DEM,
the less likely it is that the inversion will correctly determine the DEM peak
temperature (this is also shown by the tests described in section
\DUrole{ref}{modeltests}. With a narrow assumed width, plasmas which do have narrow
DEMs will at least be correctly identified, whereas plasmas with a wide DEM
would not necessarily be correctly identified by using a model DEM with a
similar width. A Gaussian with a width of \textasciitilde{}0.1 is the narrowest multi-thermal
distribution which can be distinguished from an isothermal plasma \cite{Judge2010},
so a narrower distribution would not necessarily provide meaningful results.

A Fortran extension to the main code was written to iterate through each
DEM peak temperature value for each pixel in the image, and to calculate the
corresponding goodness-of-fit value. Since the images used are very large (six
4096 x 4096 images for each temperature map), only the running best fit value
and the corresponding temperature are stored for each pixel. The temperatures
which best reproduce the observations (i.e., the temperatures with the lowest
goodness-of-fit values in each pixel) are returned to the main Python code.
Although the DEM inherently describes a multi-thermal distribution,
only the temperature of the peak of the DEM is stored and displayed in the
temperature maps. This value is useful as it is the temperature which
corresponds to the bulk temperature, and expressing the DEM as a single value
also aids visualisation.

The DEM peak temperatures considered ranged from $\log T = 5.6 - 7.0$, in
increments of 0.01 in log temperature. Outside this range of temperatures, AIA
has significantly lower temperature response and cannot provide meaningful
results. Within this range, however, the temperature is well constrained by the
response functions of the AIA channels \cite{Guennou2012} and can in principle be
calculated with a precision of \textasciitilde{}0.015 in log(T) \cite{Judge2010}.

This method is very similar in principle to the Gaussian fitting methods used
by \cite{Warren2008} and \cite{Aschwanden2013}. However, great computational efficiency
is achieved by only varying one parameter (the bulk temperature). Since the
height and width of the DEM are not investigated, this method may be less
accurate than a full parameter search would be and does not provide a full DEM
which could be used to estimate the emission measure. The width and height of
the Gaussian would need to be taken into account for a more formal
determination of the thermal structure, but this approach aims only to estimate
the dominant temperature along the LOS. The introduction of a full parameter
search will be investigated in a future work by comparing the temperature maps
produced using this implementation with those of a multi-parameter version. The
simpler implementation means that full AIA resolution temperature maps (4096 x
4096 pixels) can be calculated within \textasciitilde{}2 minutes. This is extremely fast when
compared to, for example, the multi-Gaussian fitting method used by
\cite{DelZanna2013} (which took \textasciitilde{}40 minutes to compute temperatures for 9600 pixels),
and even beats the fast DEM inversion of \cite{Plowman2012} (estimated \textasciitilde{}1 hour for
a full AIA-resolution temperature map) by a significant margin.

\subsection{Software features%
  \label{software-features}%
}

The method presented in this work stores the temperature maps as instances of
SunPy's Map object. As such, temperature maps can easily be manipulated using
any of the Map methods. For example, a temperature map of the full solar disk
can be cropped using Map.submap() in order to focus on a smaller region of the
image. The Map.plot() method also makes displaying the temperature maps very
easy.

Another advantage to using SunPy for this work is that SunPy's abilities to
query online databases makes it very easy to get AIA data and to search for
events and regions worth investigating.

The method is also able to 'track' regions over time. Since the object returned
by a database query for solar regions or events usually contains coordinate
information, those coordinates can be given to the temperature map method as a
central point around which to display the temperatures. Since the motion of
solar features is usually only dependent on the rotation of the Sun, these
features can be given a single pair of coordinates which will describe the
location of the region at any time using the Carrington Heliographic coordinate
system (which rotates at the same rate as the Sun). Therefore, any feature
can easily be 'tracked' across the Sun by this method by repeatly mapping
around these coordinates.

\section{Validation using synthetic data%
  \label{validation-using-synthetic-data}%
}

\DUrole{label}{modeltests}

Given the non-uniform nature of the instrument temperature response functions
and the \textquotedbl{}smoothing\textquotedbl{} effect of the integral equations, the accuracy of any DEM
solution will not necessarily be the same for all plasma DEMs. For instance, if
the plasma has a wide temperature distribution, the inverted DEM is less likely
to correctly identify the peak temperature than if the plasma is isothermal,
due to a reduced dependence of the DEM function on temperature \cite{Guennou2012a}.
It is therefore important to quantify the accuracy of DEM solutions with
respect to different plasma conditions as well as looking at the performance of
the method overall.

To achieve this, the method was tested by using a variety of model Gaussian
DEMs to create synthetic AIA emission, which was used as the input to the
method. The peak temperature of the model DEMs varied between 4.6 and 7.4 in
increments of 0.005, the width varied from 0.01 to 0.6 in increments of 0.005,
and the height was set at values of 15, 25 and 35. Values outside the range
scanned by the method were used in order to investigate how such values would
manifest in the temperature maps should they be present in the corona.
Similarly, the peak temperatures of the model DEMs have reduced spacing
relative to the resolution of the method in order to determine the effect
this has on the output. Only Gaussian model DEMs were used because different
multi-thermal distributions are difficult to distinguish using only AIA data
\cite{Guennou2012a} and other such shapes would therefore likely be reproduced with
similar accuracy to Gaussian DEMs. Gaussians were therefore used for
consistancy with the method itself. In any case, a full comparison of different
forms of DEM is beyond the scope of this study.

Attempting to reconstruct known DEM functions also makes it possible to
directly compare the input and output DEM functions, which is of course not
possible when using real observations. This allows a better assessment of the
accuracy of the inversions.

Figure \DUrole{ref}{model-wid001} demonstrates the accuracy of the temperature map
method when used to find model DEMs from synthesised emission. For a range of
model DEM peak temperatures and Gaussian widths and a fixed height, the plot
shows (from left to right), the peak DEM temperature inferred by the method,
the percentage diference between the solution and the true DEM peak temperature,
and the goodness-of-fit values associated with the solutions. The temperatures
obtained using this method vary only with the peak temperature and width of the
model DEM; varying the height of the model DEM appears to cause no change in
the solution.

For model DEM widths of < 0.1, model DEM peak temperatures within the range
considered by the temperature map method are generally found with reasonable
accuracy, and with similar accuracy for all temperatures in this range apart
from a sharp drop in solution temperature at a model DEM temperature of
log(T) = 6.4 - 6.45. Hotter model DEMs are also fairly well matched as they
produce solution temperatures of log(T) $\approx$ 7.0, though the
solution temperature drops off slightly as the model DEM peak temperature
increases, reducing the accuracy. Cooler model DEMs are less well reproduced
by the method, with the solution increasing as the model peak temperature
decreases down to log(T) $\approx$ 5.1, and falling again thereafter.
The goodness-of-fit values are lowest for model DEM peaks between log(T) = 5.6
and $\approx$ 6.1, and generally increase for temperatures above this
range, whereas they are relatively low at cooler temperatures.

The results are significantly better for model DEMs with a width of 0.1, which
is equal to the width assumed by the method. Model temperatures within the
range of the method are reproduced almost exactly and with goodness-of-fit
values $\ll$ 1 in most cases. Again, the solution temperature drops with
increasing model temperature above log(T) = 7.0. Below log(T) = 5.6, however,
the method returns a temperature of log(T) $\approx$ 6.1 for all model
temperatures. Goodness-of-fit values at temperatures above and below the
method's range are relatively low (\textasciitilde{}0.01 - 1.0), with those at higher
temperatures being larger.

In the case of much wider model DEMs (> 0.45) the solution temperature has no
dependence at all on the model peak temperature, and returns log(T)
$\approx$ 6.1 for all model DEMs. However, the goodness-of-fit values are
still quite low ($<$ 0.01) for all model DEMs despite the significant
failure of the method for these conditions.\begin{figure*}[]\noindent\makebox[\textwidth][c]{\includegraphics[scale=0.35]{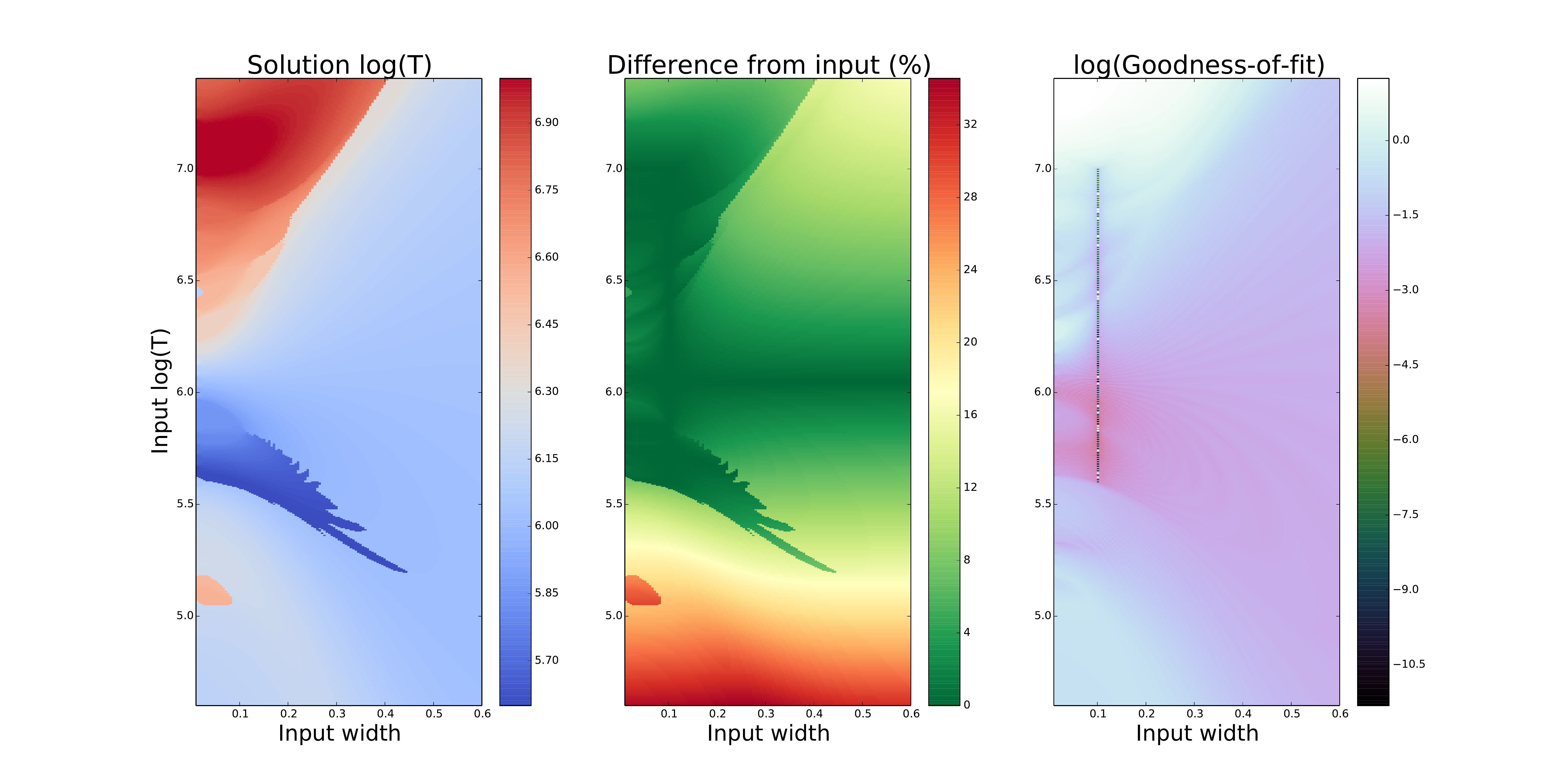}}
\caption{Assessment of method accuracy for model DEMs with various peak temperatures
and widths, and with a constant height. Results for all values of emission
measure tested were found to be identical. Left: peak DEM temperature found
by method. Middle: absolute difference between solution and model DEM peak
temperature as a percentage of the latter. Right: goodness-of-fit values
corresponding to solution temperatures, shown on a logarithmic scale. Lower
values indicate a better fit to the observations. The values of the left,
middle and right plots are shown for a DEM width of 0.1 in Figures
\DUrole{ref}{solution-slice}, \DUrole{ref}{diff-slice} and \DUrole{ref}{fit-slice}, respectively.
\DUrole{label}{model-wid001}}
\end{figure*}\begin{figure}[]\noindent\makebox[\columnwidth][c]{\includegraphics[width=\columnwidth]{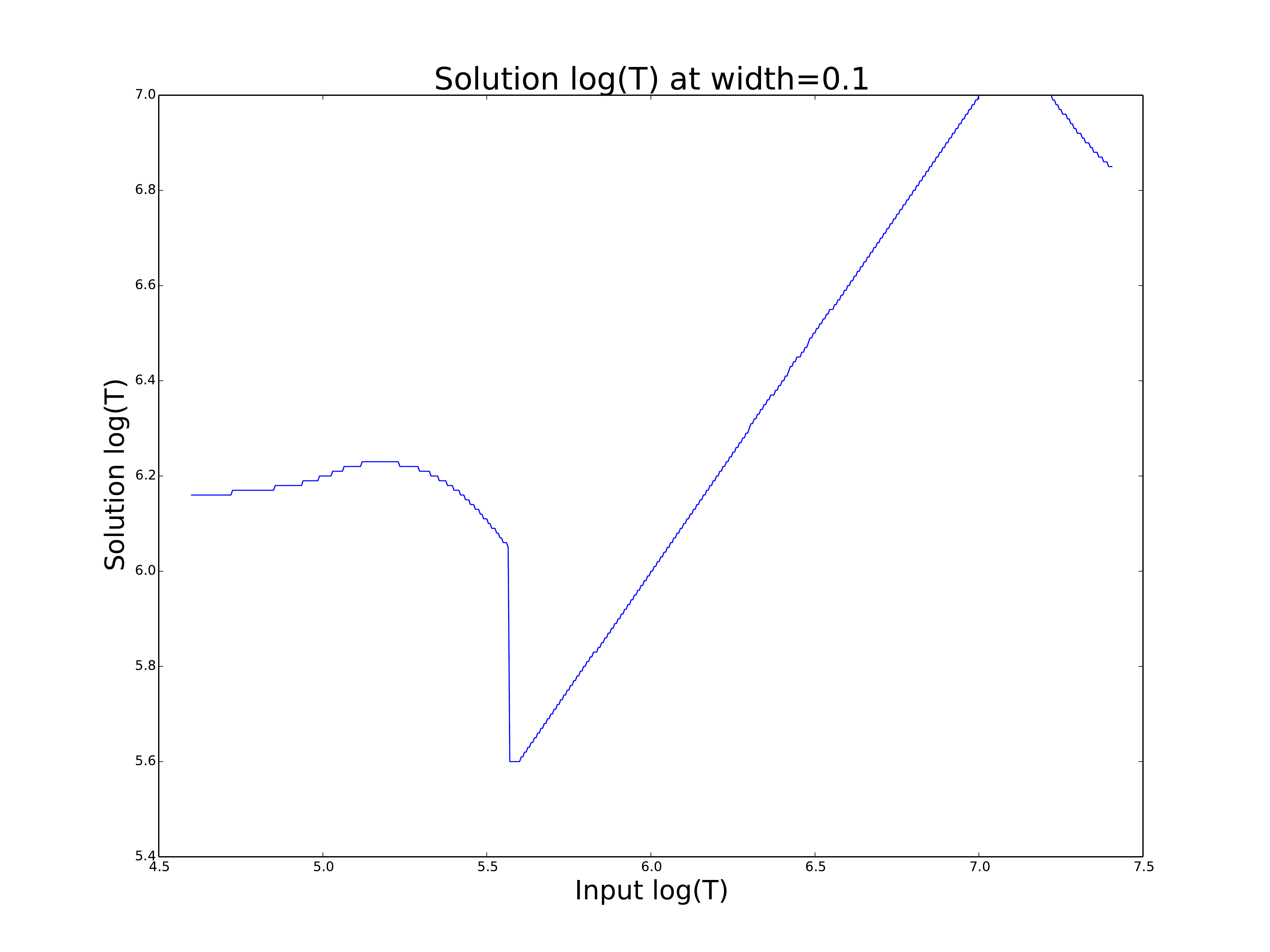}}
\caption{Plot of the solution temperature produced for a given input DEM peak
temperature and a DEM width of 0.1. \DUrole{label}{solution-slice}}
\end{figure}\begin{figure}[]\noindent\makebox[\columnwidth][c]{\includegraphics[width=\columnwidth]{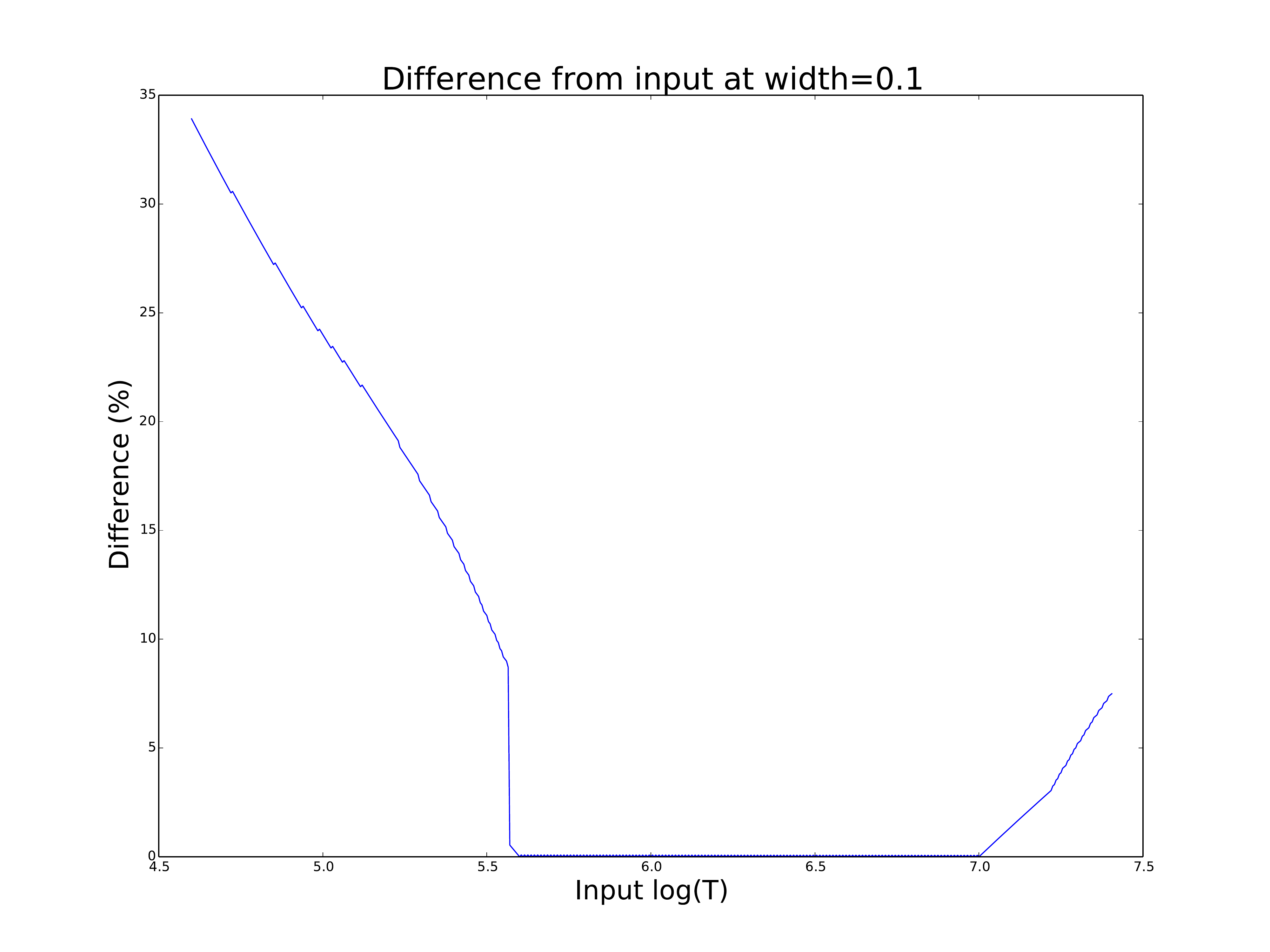}}
\caption{Plot of the percentage difference between input and solution temperatures
for a DEM width of 0.1. \DUrole{label}{diff-slice}}
\end{figure}\begin{figure}[]\noindent\makebox[\columnwidth][c]{\includegraphics[width=\columnwidth]{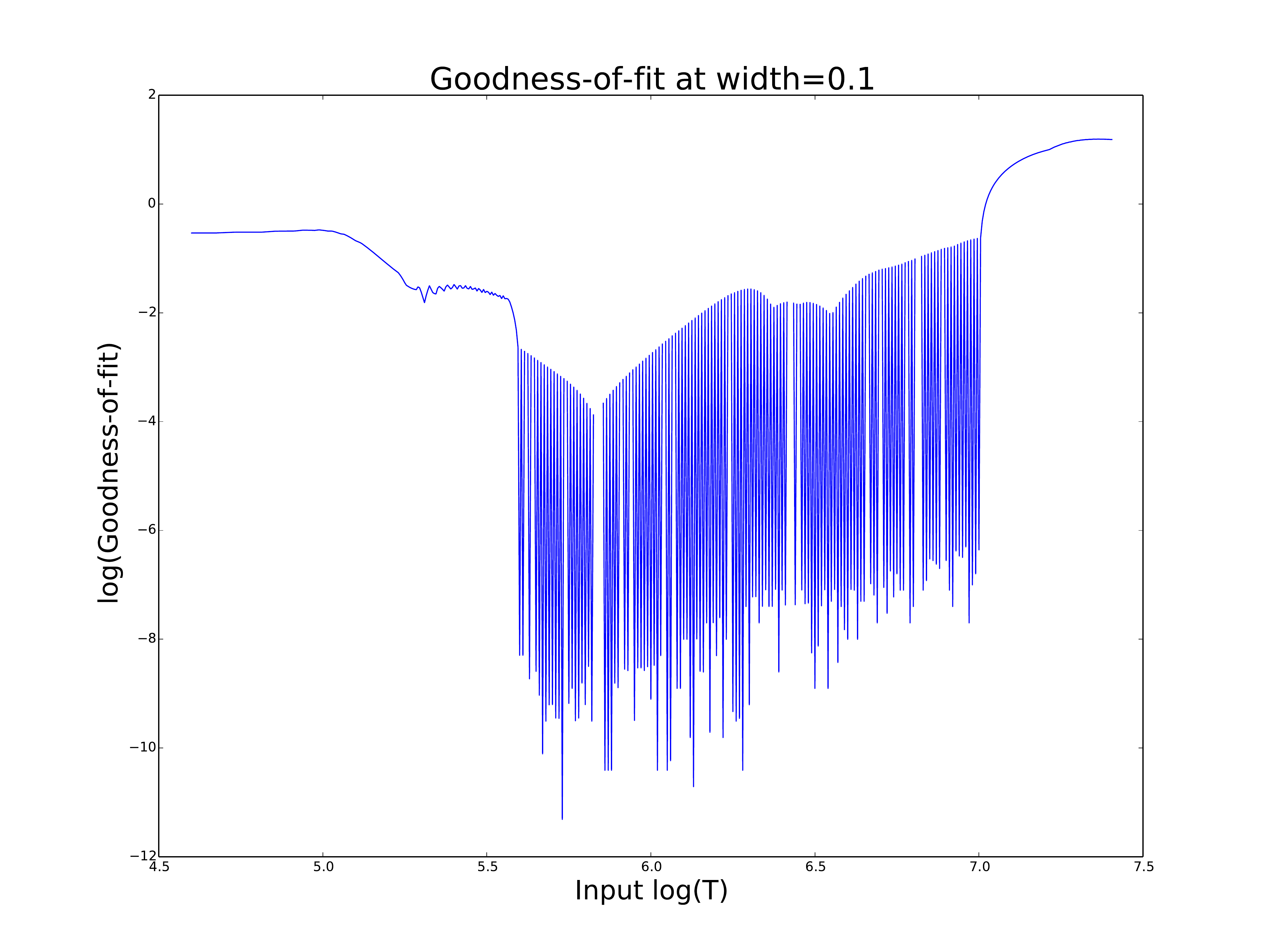}}
\caption{Plot of the goodness-of-fit values produced for a given input DEM peak
temperature and a DEM width of 0.1. \DUrole{label}{fit-slice}}
\end{figure}

\section{Results%
  \label{results}%
}

\DUrole{label}{results}

The temperature maps calculated using the proposed method and the method
described in \cite{Aschwanden2013} are shown in Figures \DUrole{ref}{mytemps} and
\DUrole{ref}{aschtemps} respectively. The Aschwanden method is used for this
comparison because it is recent and similar to the propsed method, and because
few other papers present full-disk temperature maps. For ease of comparison,
the results of this work are plotted using a similar colour map to the one used
by \cite{Aschwanden2013} and with the same upper and lower temperature limits.

The two methods find similar temperatures for the majority of the corona,
though regions found to have extreme hot or cool temperatures using
Aschwanden's method were closer to average in the map calculated with the
proposed method. Also note that Figure \DUrole{ref}{mytemps} was calculated using
full-resolution AIA data, whereas Aschwanden's method rebins the original data
into 4x4 macropixels (i.e. 1024x1024 images).\begin{figure}[]\noindent\makebox[\columnwidth][c]{\includegraphics[width=\columnwidth]{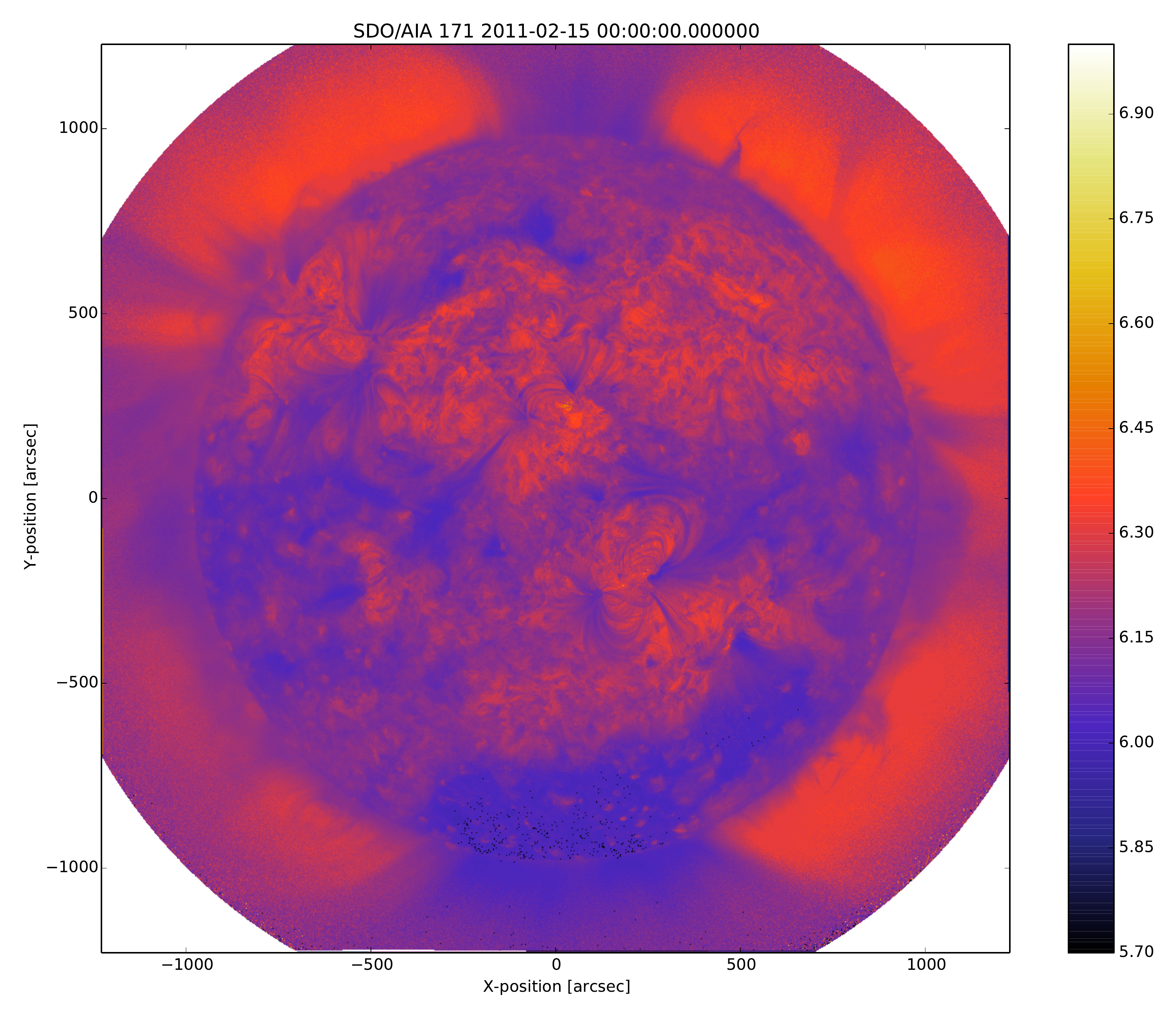}}
\caption{Temperature map for the full-disk corona on 2011-02-15 00:00, calculated
using the proposed method. The colour map and temperature limits were
chosen to match those in Figure \DUrole{ref}{aschtemps}. The X-position and
Y-position of the axis refer to arc seconds from solar disk centre in the
observer's frame of reference, with the Y-position aligned to solar north.
\DUrole{label}{mytemps}}
\end{figure}\begin{figure}[]\noindent\makebox[\columnwidth][c]{\includegraphics[width=\columnwidth]{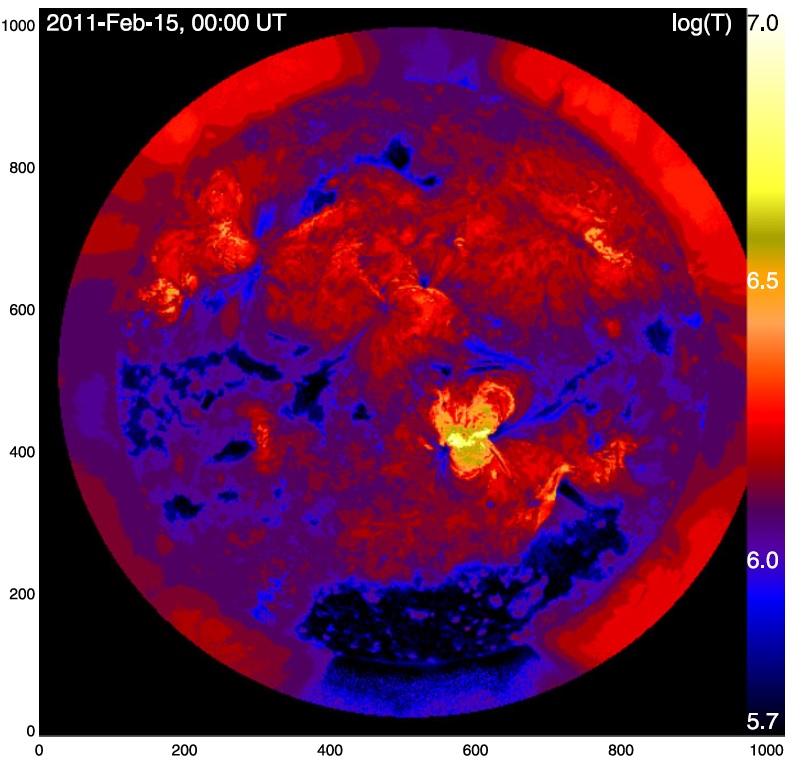}}
\caption{Temperature map for the full-disk corona on 2011-02-15 00:00. Image from
\cite{Aschwanden2013}. \DUrole{label}{aschtemps}}
\end{figure}

The remaining results have been sectioned into three general regions of the
corona - quiet sun, coronal holes and active regions. All regions studied were
selected from January and February 2011.

\subsection{Quiet sun%
  \label{quiet-sun}%
}

The term 'quiet sun' refers to the portions of the Sun in which there is little
or no activity. In many cases this will be the majority of the solar disk.
Three large regions of quiet sun were selected. The criterion for selection was
simply a nondescript region of the disk near disk centre not containing active
regions, coronal holes or dynamic events (e.g. coronal jets). Figures
\DUrole{ref}{qs20110128}, \DUrole{ref}{qs20110208} and \DUrole{ref}{qs20110221} show three regions
on 2011-01-28 00:00, 2011-02-08 00:00 and 2011-02-21 00:00 respectively (these
figures have all been plotted to the same colour scale for ease of comparison).
The quiet sun regions on 2011-01-28 and 2011-02-08 were found to have very
similar temperature distributions, with minima of log(T) = 5.97 and 5.99, means
of log(T) = 6.08 and 6.09, and maxima of log(T) = 6.31 and 6.31 respectively.
The temperature map for 2011-02-21 found mostly similar temperatures to the
previous two regions, apart from a few isolated pixels with spurious values.
The mean for this region was log(T) = 6.08. The minimum value, excluding
spurious pixels, is log(T) = 5.96 and the maximum is log(T) = 6.29.

In all three temperature maps the hottest temperatures are found in relatively
small, localised regions (which appear in red in Figures \DUrole{ref}{qs20110128} and
\DUrole{ref}{qs20110208}), with the temperatures changing quite sharply between these
regions and the cooler background plasma. These hotter regions appear to
consist of small loop-like structures, though none of these correspond to any
active region (see section \DUrole{ref}{ARs}). The hot structures in the region shown
in Figure \DUrole{ref}{qs20110221} take up a slightly larger portion of the region and
are more strongly concentrated in one location. Temperatures of around log(T)
$\approx$ 6.15 also appear to form even smaller loops in some cases,
which are more evenly distributed than the hotter regions. Temperatures below
this are more uniform and have no clearly visible structure.\begin{figure}[]\noindent\makebox[\columnwidth][c]{\includegraphics[width=\columnwidth]{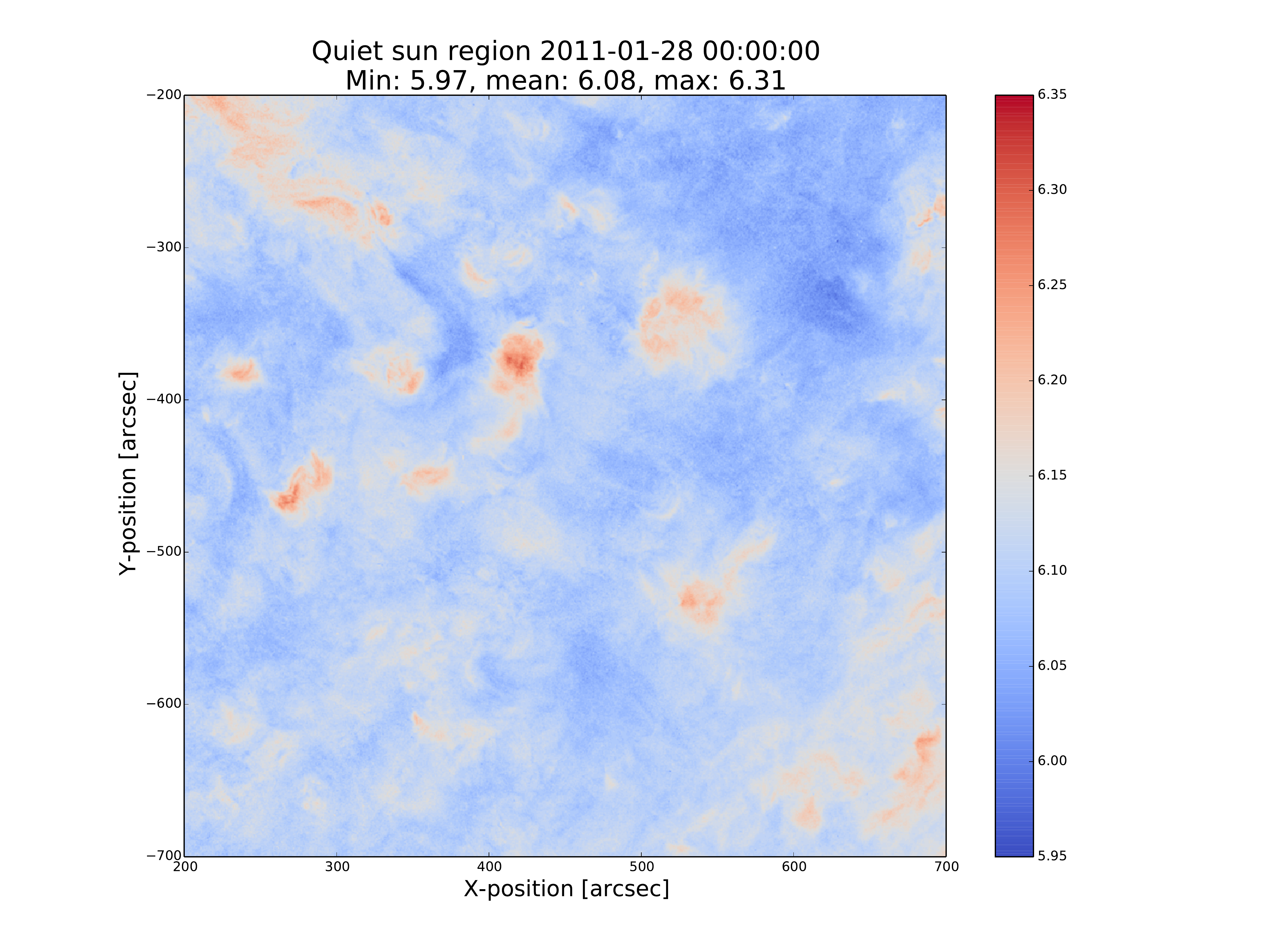}}
\caption{\DUrole{label}{qs20110128}
Temperature map for quiet sun region on 2011-01-28 00:00.}
\end{figure}\begin{figure}[]\noindent\makebox[\columnwidth][c]{\includegraphics[width=\columnwidth]{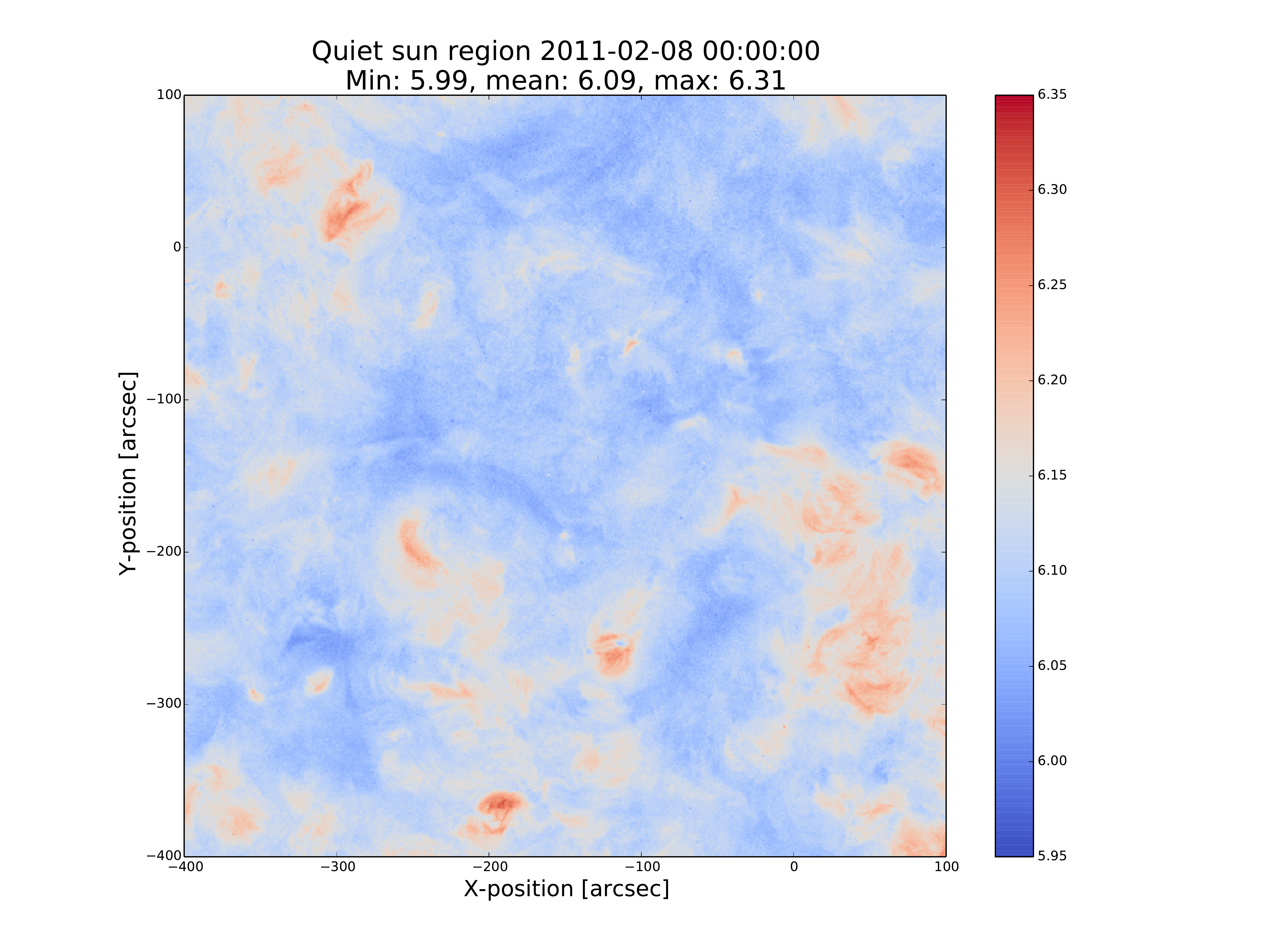}}
\caption{\DUrole{label}{qs20110208}
Temperature map for quiet sun region on 2011-02-08 00:00.}
\end{figure}\begin{figure}[]\noindent\makebox[\columnwidth][c]{\includegraphics[width=\columnwidth]{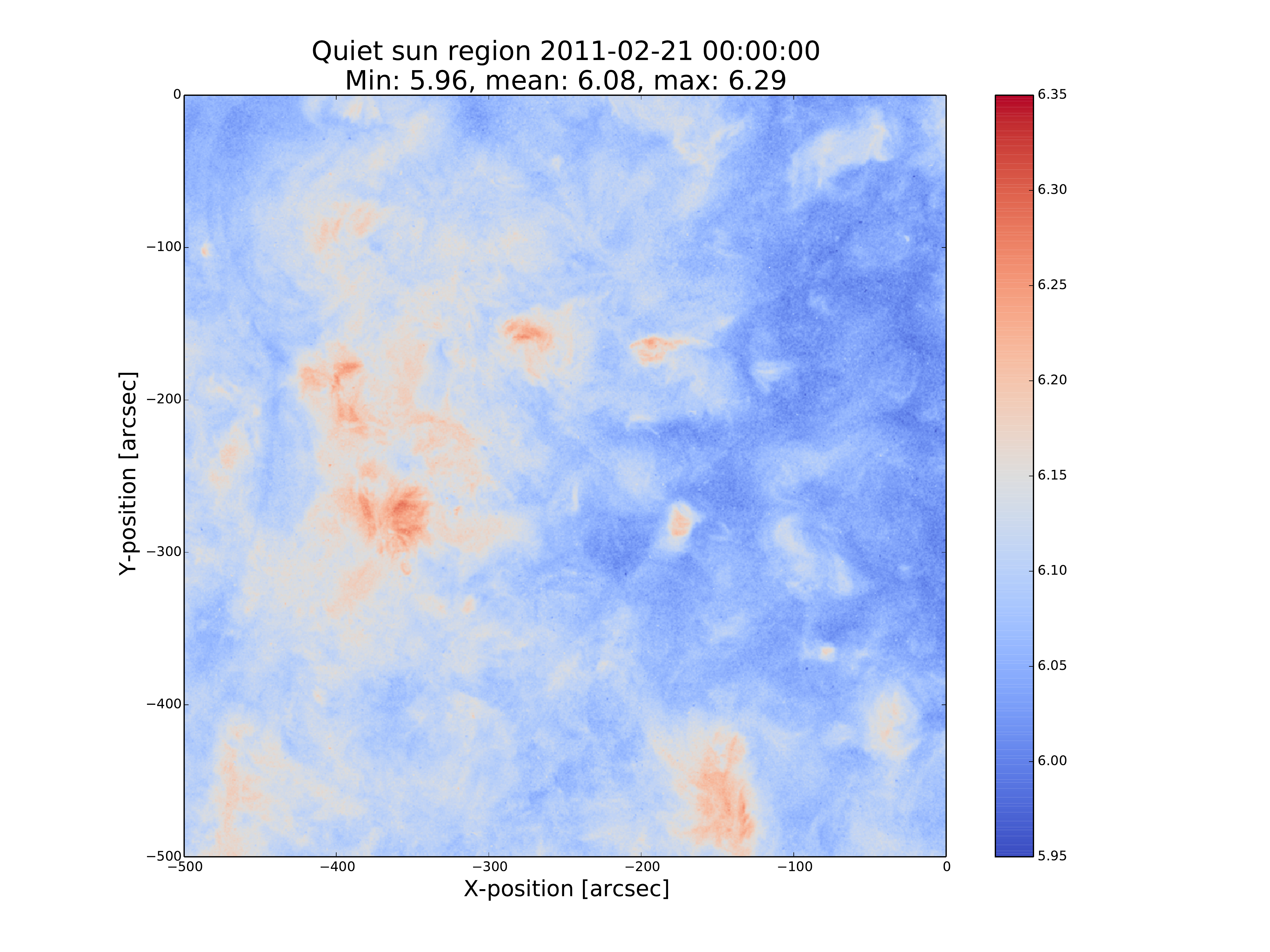}}
\caption{\DUrole{label}{qs20110221}
Temperature map for quiet sun region on 2011-02-21 00:00. Spurious
low-temperature pixels have been removed.}
\end{figure}

\subsection{Coronal holes%
  \label{coronal-holes}%
}

Coronal holes are regions of effectively open magnetic field which exhibit very
low levels of emission in EUV and X-ray wavelengths. Figures \DUrole{ref}{ch20110201a},
\DUrole{ref}{ch20110201b} and \DUrole{ref}{ch20110214} show temperature maps for coronal
holes. Note that these figures are shown with different colour scales to each
other and to figures \DUrole{ref}{qs20110128}, \DUrole{ref}{qs20110208} and \DUrole{ref}{qs20110221}.
In Figures \DUrole{ref}{ch20110201b} and \DUrole{ref}{ch20110214} the solar limb (the edge of
the disk of the Sun) is marked with a black line. Figure \DUrole{ref}{limbtemps} shows
the temperatures along the vertical line shown in Figure \DUrole{ref}{ch20110214}. The
coronal holes shown in figures \DUrole{ref}{ch20110201a} and \DUrole{ref}{ch20110201b}
(henceforth coronal holes 1 and 2), were observed at 2011-02-01 00:00 in the
northern and southern hemispheres respectively, and the one in figure
\DUrole{ref}{ch20110214} (coronal hole 3) was observed at 2011-02-14 00:00. The
minimum, mean and maximum temperatures found for the regions mapped were:
log(T) = 5.6, 6.03 and 6.52 for coronal hole 1; log(T) = 5.6, 6.02 and 6.32 for
coronal hole 2; and log(T) = 5.6, 6.02 and 6.37 for coronal hole 3. The
somewhat higher maximum temperature for coronal hole 1 appears to be due to
hotter material above the solar limb over the quiet sun regions. Such
unavoidable contamination of the coronal hole data by other non-coronal hole
structures along the line of sight can, in principle, be reduced using
tomographical reconstruction techniques such as the one described by
\cite{Kramar2014}.

In all three figures, the coronal hole region is clearly visible as a region of
significantly cooler plasma than the surrounding quiet sun regions, with the
former mostly exhibiting temperatures in the range log(T) $\approx$
5.9 - 6.05, and the latter being mostly above log(T) $\approx$ 6.1. In
all three coronal holes, though to a much greater extent in coronal holes 2 and
3, a 'speckling' effect is observed, which is caused by numerous very small low
temperature regions. Each of these consists only of a few pixels and were found
to have temperatures of log(T) $\approx$ 5.6-5.7. This speckling is
similar to the individual low-temperature pixels found for the quiet sun region
for 2011-02-21 (Figure \DUrole{ref}{qs20110221}), but is much more prominent.

All three coronal holes also contain small hotter regions (log(T) $\ge$
6.1), which appear to be similar to quiet sun regions and in some cases seem to
consist of closed loop-like structures within the larger open magnetic field of
the coronal hole. In addition to these regions, coronal hole 1 contains a large
quiet sun region.

The temperature over coronal holes 2 and 3 (i.e. below the limb in Figures
\DUrole{ref}{ch20110201b} and \DUrole{ref}{ch20110214}) was found to increase slightly with
distance from the centre of the image. This temperature gradient is plotted for
coronal hole 3 in Figure \DUrole{ref}{limbtemps}. In both cases, the temperature is
log(T) $\approx$ 6.0 at the limb and rises to log(T) $\approx$ 6.05
at the edge of the mapped region.\begin{figure}[]\noindent\makebox[\columnwidth][c]{\includegraphics[width=\columnwidth]{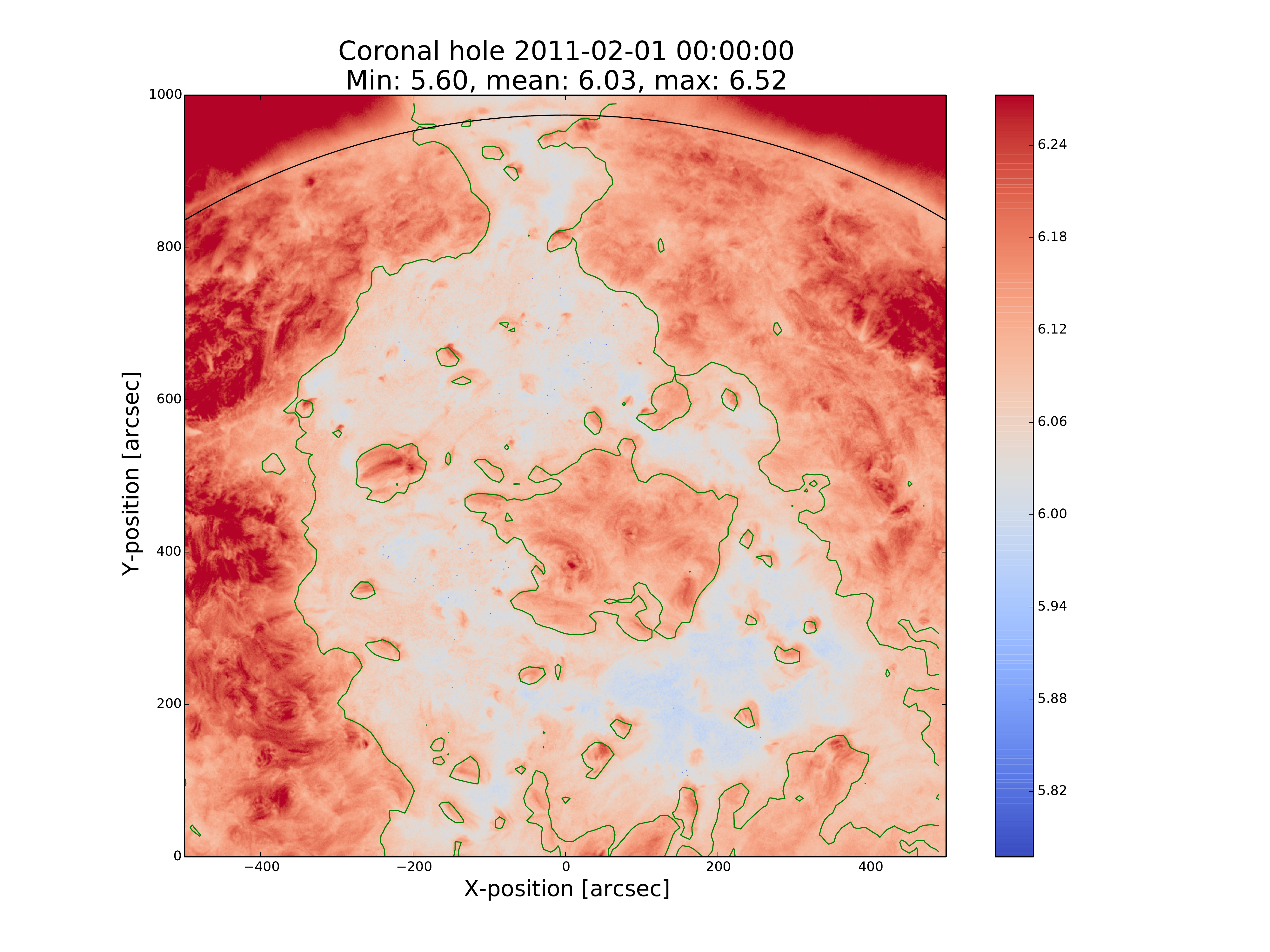}}
\caption{Temperature map of the northern coronal hole at 2011-02-01 00:00 (coronal
hole 1). The coronal hole itself is clearly visible as the blue-white
region, with the surrounding quiet sun plasma appearing in red. A few
isolated low-temperature pixels can be seen inside the boundaries of the
coronal hole, as well as a large quiet sun region and several smaller ones.
\DUrole{label}{ch20110201a}}
\end{figure}\begin{figure}[]\noindent\makebox[\columnwidth][c]{\includegraphics[width=\columnwidth]{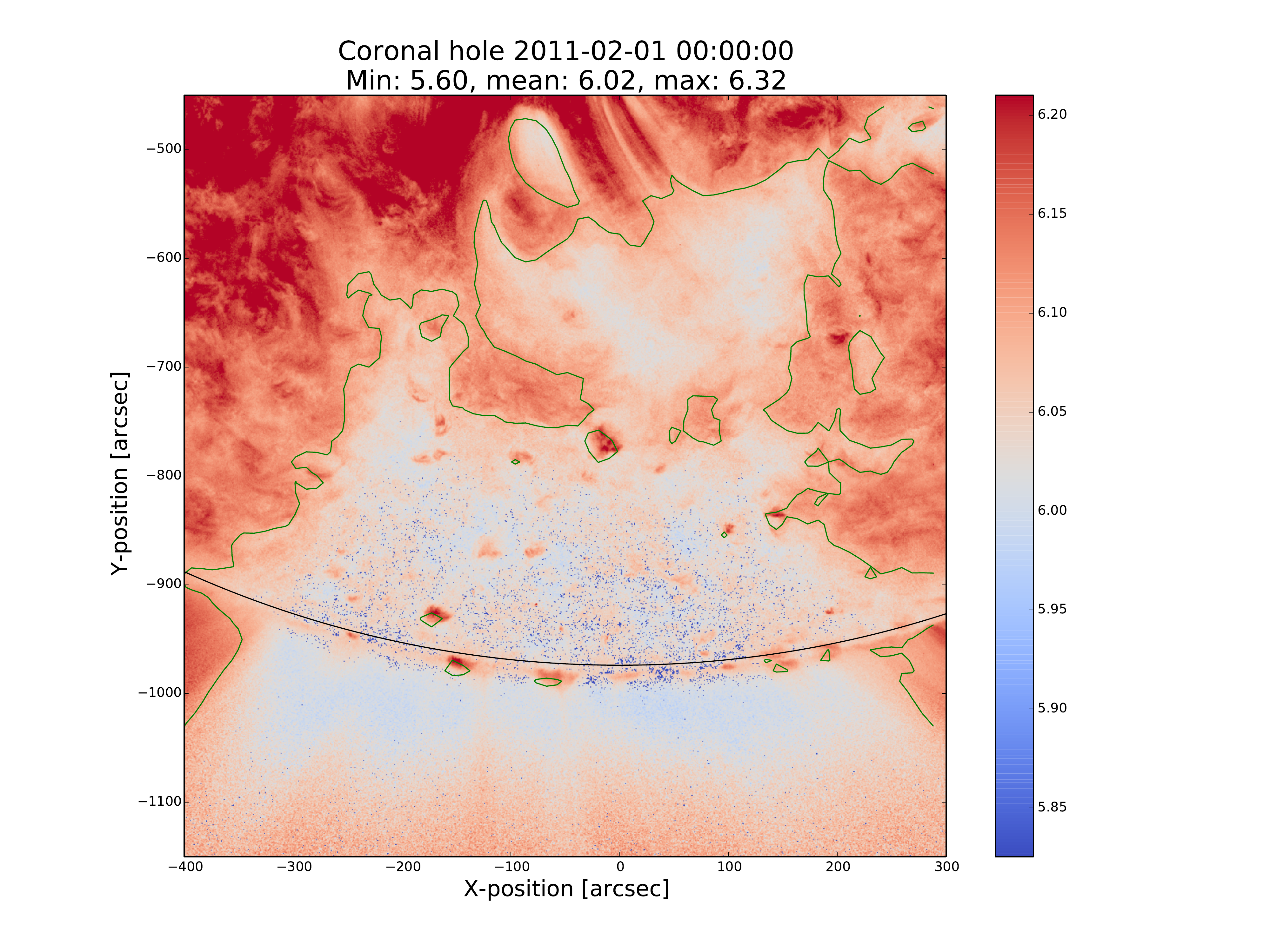}}
\caption{Temperature map of the southern coronal hole at 2011-02-01 00:00 (coronal
hole 2). As with Figure \DUrole{ref}{ch20110201a}, the coronal hole stands out
against the hotter quiet sun. This region shows much more 'speckling'
within the coronal hole from low-temperature pixels, but contains several
small quiet sun like regions similar to those seen in coronal hole 1.
\DUrole{label}{ch20110201b}}
\end{figure}\begin{figure}[]\noindent\makebox[\columnwidth][c]{\includegraphics[width=\columnwidth]{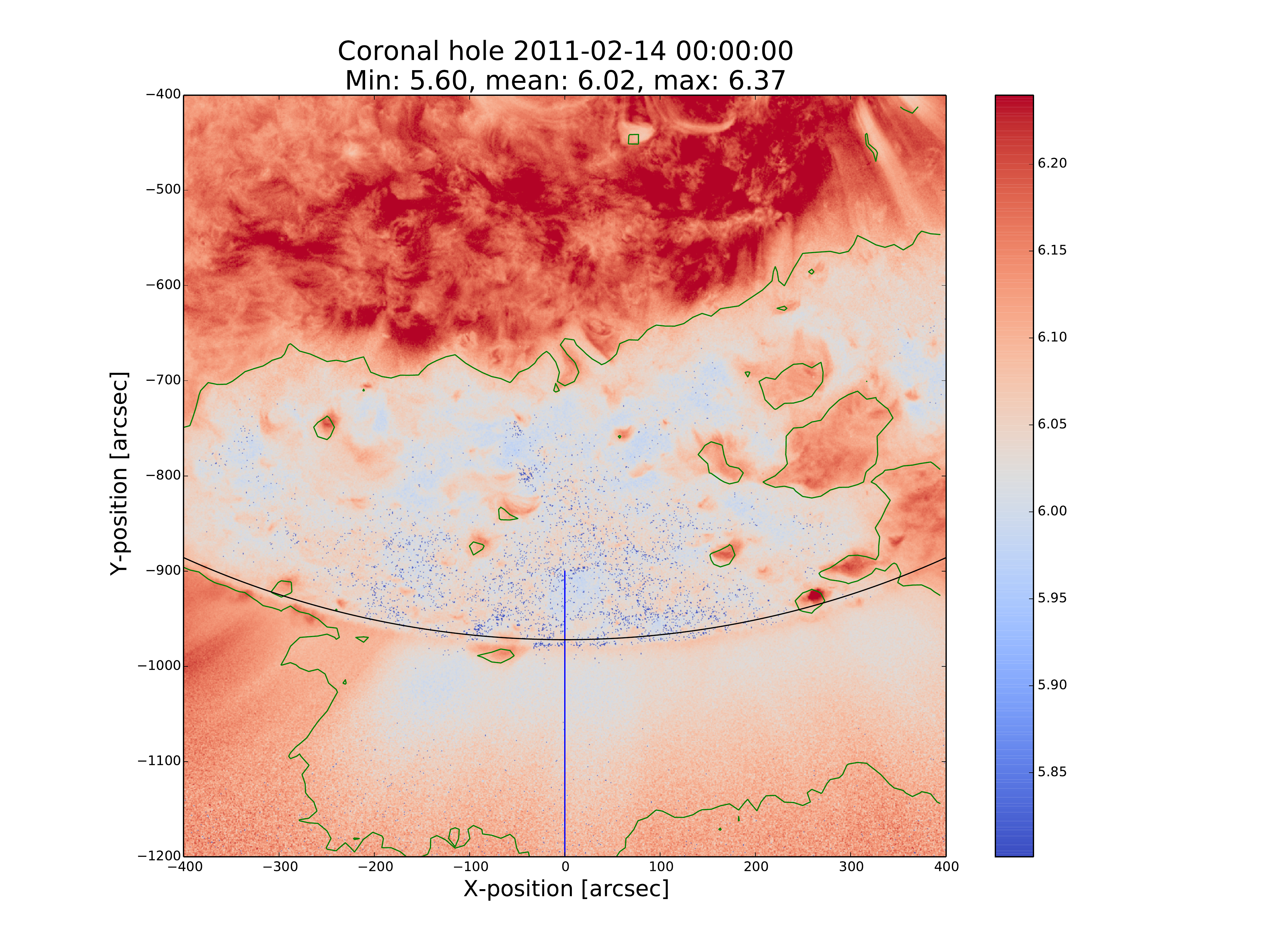}}
\caption{Temperature map of a coronal hole at 2011-02-14 00:00 (coronal hole 3).
Again, this coronal hole contains significant speckling and several small
quiet sun like regions. \DUrole{label}{ch20110214}}
\end{figure}\begin{figure}[]\noindent\makebox[\columnwidth][c]{\includegraphics[width=\columnwidth]{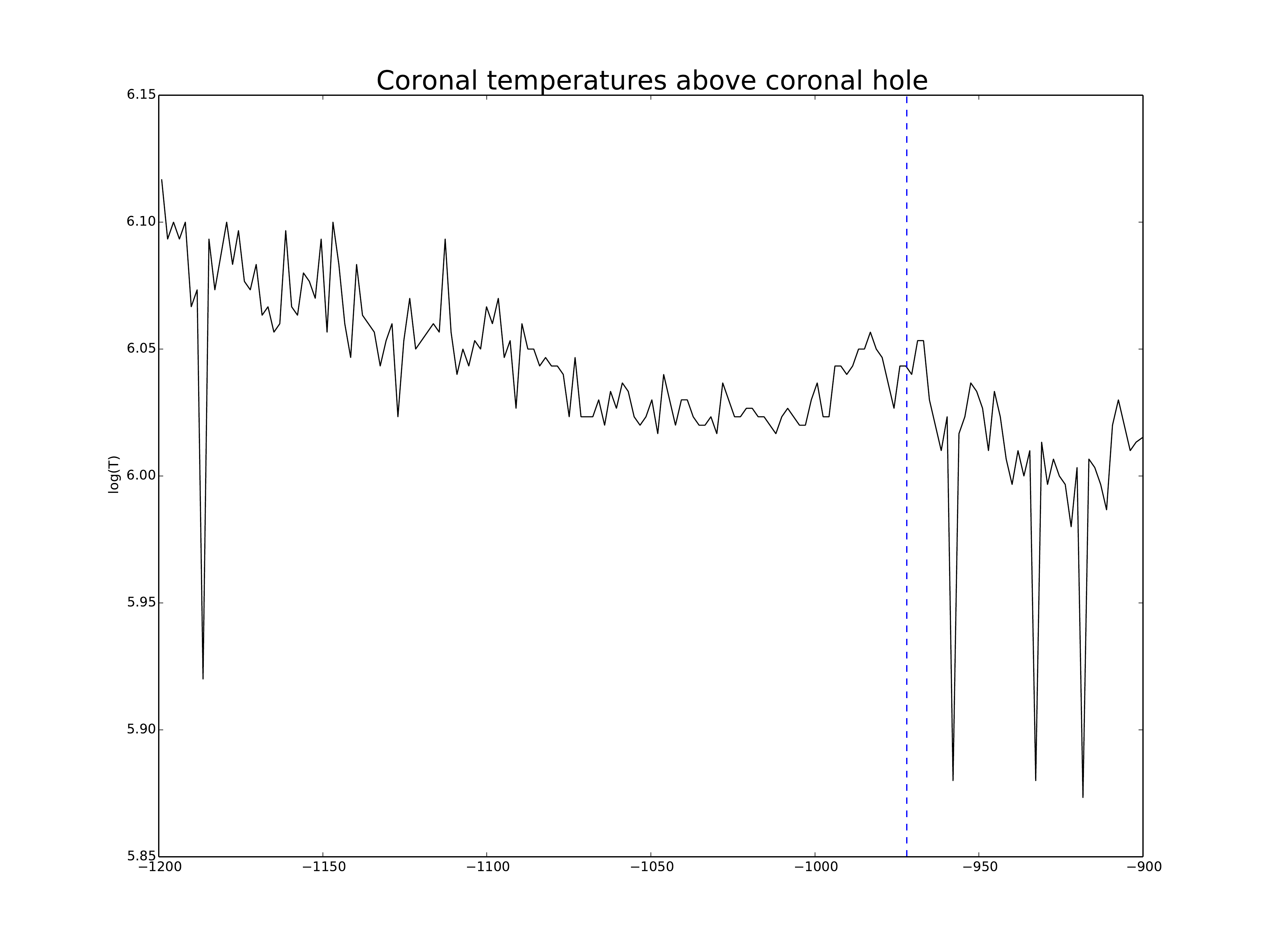}}
\caption{Plot of temperatures along the vertical line show in Figure
\DUrole{ref}{ch20110214}. Note that these temperatures have been smoothed to more
clearly show the overall trend. \DUrole{label}{limbtemps}}
\end{figure}

\subsection{Active regions%
  \label{active-regions}%
}

\DUrole{label}{ARs}

Active regions are areas of concentrated magnetic field, and consist of many
magnetic field lines ('coronal loops', or often simply 'loops') which are
seen in EUV and X-rays as strands of bright material. The points at which these
loops are rooted in the lower corona are called footpoints.

Active regions show the greatest variation in temperature, as can be seen in
Figures \DUrole{ref}{ar20110122}, \DUrole{ref}{ar20110201} and \DUrole{ref}{ar20110219}. These
figures show temperature maps of active regions AR11147 and AR11149 (henceforth
region 1), active region AR11150 (region 2) and active regions AR11161 and
AR11162 (region 3), respectively. Regions 1 and 3 are much more complex than
region 2, as each consists of a larger main active region and a smaller region
which has emerged nearby. The minimum, mean and maximum temperatures found were:
log(T) = 6.03, 6.2 and 6.54 for region 1; log(T) = 6.05, 6.22 and 6.41 for
region 2; and log(T) = 6.01, 6.22 and 6.57 for region 3.

In each of these regions, the coolest temperatures are found in the largest
loops with footpoints at the edges of the active region, which were found to
have temperatures between log(T) = 6.05 and log(T) = 6.1. Smaller loops with
footpoints closer to the centre of the active region show higher temperatures
(log(T) $\approx$ 6.1 - 6.3). Hotter temperatures again (log(T)
$\ge$ 6.3) were also found in all three active regions, though in
different locations. In region 1 these temperatures can be seen in parts of the
very small loops in AR11149, as well as in what may be small loops or
background in AR11147. In region 2 they are found in loops which appear to be
outside the main active region. In region 3 they are found in a few relatively
large loops - in contrast to the much smaller loops found to have those
temperatures in the other regions - and there are also several small hot
regions around AR11162 near the top of Figure \DUrole{ref}{ar20110219}. The noisy
nature of the hot regions in this figure appears to be due to unusually high
relative values in the typically noisy 9.4nm and 13.1nm channels, which
correspond to high temperature plasma.

All three regions also show the presence of cooler quiet sun-like plasma
surrounding the active regions (log(T) $\approx$ 6.1 - 6.2), and Figure
\DUrole{ref}{ar20110201} shows a filament found to have a fairly uniform temperature
of log(T) $\approx$ 6.3.\begin{figure}[]\noindent\makebox[\columnwidth][c]{\includegraphics[width=\columnwidth]{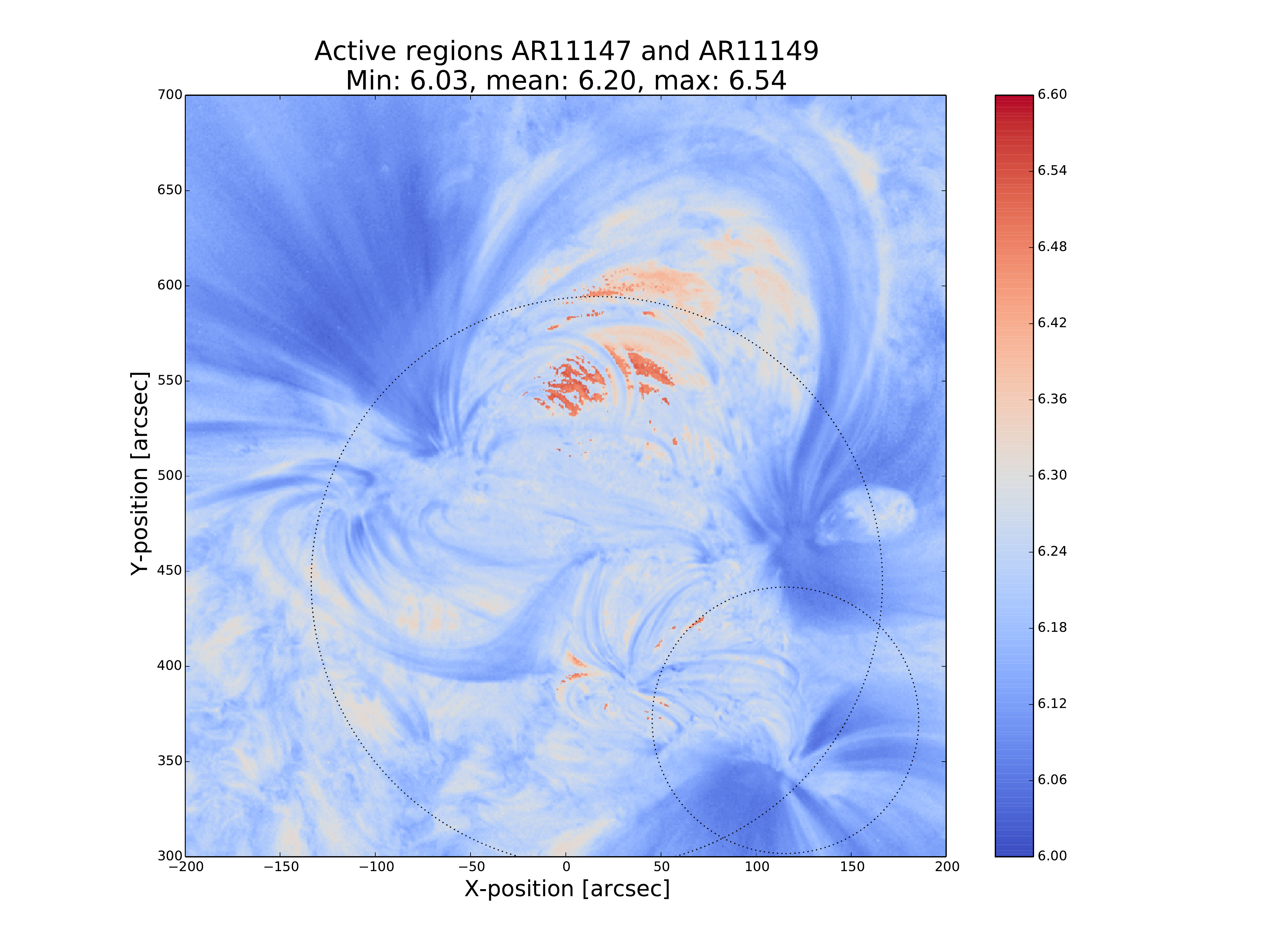}}
\caption{Temperature map of active region AR11147 and AR11149 at 2011-01-22 00:00.
The large and small circles indicate the general areas of AR11147 and
AR11149 respectively. Large, cool loops appear in dark blue, with loop
temperatures generally increasing as the loop size decreases. Quiet
sun-like plasma is also visible around the active region in shades of light
blue. \DUrole{label}{ar20110122}}
\end{figure}\begin{figure}[]\noindent\makebox[\columnwidth][c]{\includegraphics[width=\columnwidth]{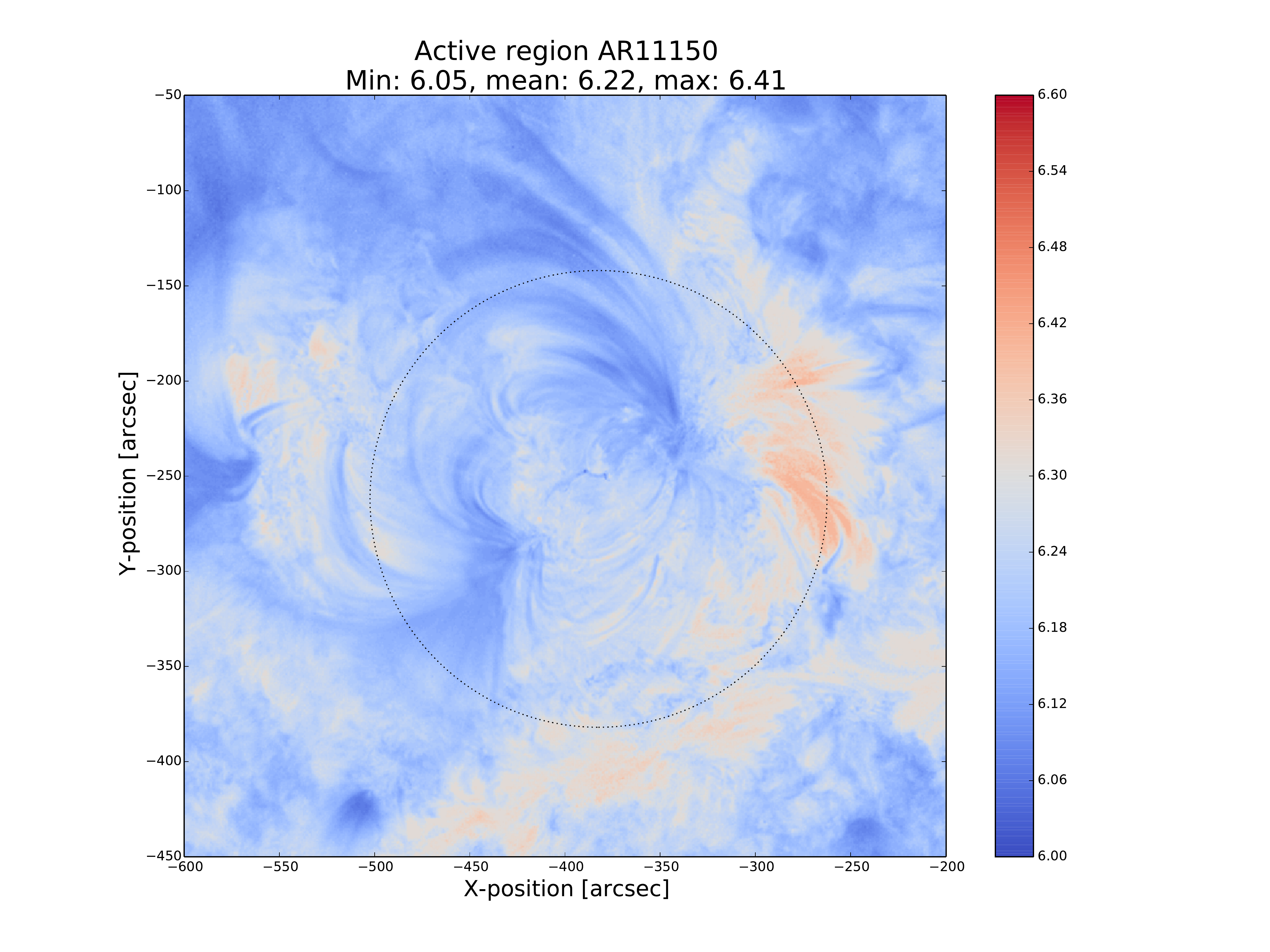}}
\caption{Temperature map of active region AR11147 at 2011-02-01 00:00. The circle
indicates the general area of the active region. As in figure
\DUrole{ref}{ar20110122}, the largest loops exhibit the lowest temperatures and
cool quiet sun plasma surrounds the region. Also seen is a filament with a
roughly uniform temperature of log(T) \textasciitilde{}6.3. \DUrole{label}{ar20110201}}
\end{figure}\begin{figure}[]\noindent\makebox[\columnwidth][c]{\includegraphics[width=\columnwidth]{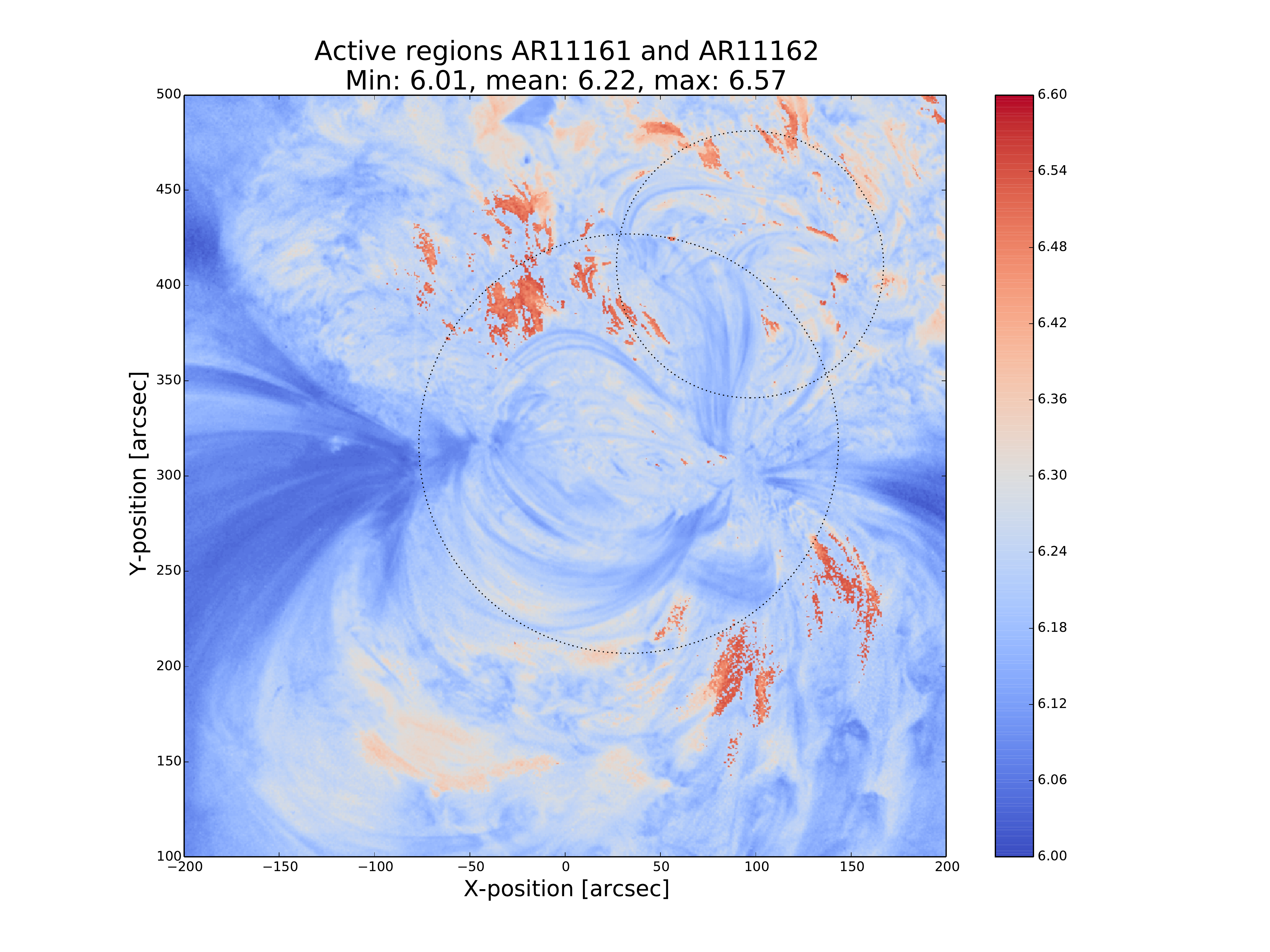}}
\caption{Temperature map of active region AR11161 and AR 11162 at 2011-02-19 00:00.
The large and small circles indicate the general areas of AR11161 and
AR11162 respectively. Again, low temperatures are found in large loops and
quiet sun like plasma is seen around AR11161. However, unlike regions 1
and 2, the hottest temperatures here are found in some relatively large
loops and in small patches around AR11162 where one would expect to find
much cooler plasma. \DUrole{label}{ar20110219}}
\end{figure}

\section{Discussion%
  \label{discussion}%
}

\DUrole{label}{disc}

The proposed method produces results many times faster than typical DEM methods,
with a full-resolution temperature map being produced in \textasciitilde{}2 minutes. The great
efficiency of the method makes it well suited for realtime monitoring of the
Sun. The challenge lies in finding connections between changes of temperature
with time, or between changes in the spatial distribution of temperature, with
events of interest (e.g. large flares). The realtime prediction of large events
would be a very desirable goal. This is work we are currently undertaking.
Results over the whole solar disk with reasonably high time resolution also
allows us to make statistical studies of the way temperature changes within
certain regions over long time periods. This is another approach we are
currently using to study active regions in particular.

For some quiet sun regions and coronal holes, the method found
low-temperature values for isolated pixels or for small groups of pixels.
It is possible that these isolated pixels are due to one or more channels being
dominated by noise which is amplified by the normalisation of the images.
However, these pixels are also seen far more in coronal holes 2 and 3 (Figures
\DUrole{ref}{ch20110201b} and \DUrole{ref}{ch20110214}), which were observed at the pole,
than in coronal hole 1 (Figure \DUrole{ref}{ch20110201a}), which ranged from near the
pole to near the equator. It is therefore also possible these cold pixels are
at least partly due to some LOS effect.

The temperature values found for active regions are largely as was expected,
though they are slightly cooler in places than some other studies have found.
It is important to bear in mind when considering active regions that the
assumptions on which the temperature map method depends may not be met, such as
the assumption of local thermal equilibrium. Additionally, no background
subtraction has been applied to the AIA images used, which may account for some
of the discrepancy between these results and those of other authors.

Figure \DUrole{ref}{ar20110201} includes a filament, which was found to have a fairly
uniform temperature of log(T) $\approx$ 6.3. This contradicts the
established wisdom that filaments consist of cooler plasma than much of the
rest of the corona, and probably indicates a failing of this temperature method.
Since filaments are relatively dense structures and this method does not take
into account density, it is likely that the plasma conditions found in
filaments are poorly handled by the method. This suggests it may be unwise to
rely too heavily on this method for temperatures of filaments or similarly
dense coronal structures.

As discussed in Section \DUrole{ref}{modeltests}, narrow DEMs widths are
reconstructed much more accurately than wide ones, with solutions tending
towards \textasciitilde{}1MK with increasing DEM width. Such results in these temperature maps
should therefore be treated with a certain amount of caution. Overall, however,
the temperature map method performs very well and produces temperatures which
are consistent with the results of previous studies. A slower but more complete
version which fits a full DEM to the observations will be the focus of a later
work and will provide more information on the corona's thermal structure.

An important point is that producing temperature maps across such large regions
was impossible until AIA/SDO began observations. The results presented in this
paper are therefore unique and new. Code written almost exclusively in the
Python language has been used to produce the results, and Python has been
instrumental in ensuring the efficiency of the processing. Whilst many other
groups are using AIA/SDO to estimate or constrain temperatures, our approach is
to develop the most efficient and quick code that will allow us to make large
statistical studies, studies of temporal changes, and search for predictable
connections between temperature changes and large events.

\section{Acknowledgements%
  \label{acknowledgements}%
}

This work is funded by an STFC student grant.

This research has made use of SunPy, an open-source and free
community-developed solar data analysis package written in Python
\cite{Mumford2013}.

\end{document}